\documentclass[%
 reprint,
 amsmath,amssymb,
 aps,
]{revtex4-2}

\usepackage{graphicx}
\usepackage{dcolumn}
\usepackage{bm}
\usepackage{orcidlink}
\usepackage{subfigure}
\usepackage{multirow}
\usepackage{float}

\usepackage[normalem]{ulem}

\begin{document}

\preprint{APS/123-QED}

\title{Systematic bias in dark siren statistical methods and its impact on Hubble constant measurement}

\author{Viviane Alfradique\orcidlink{0000-0002-5225-1923}}
\email{vivianeapa@cbpf.br}
 \affiliation{
 Centro Brasileiro de Pesquisas F\'isicas, Rua Dr. Xavier Sigaud 150, \\
 22290-180 Rio de Janeiro, RJ, Brazil
}%
\author{Clécio R. Bom\orcidlink{0000-0003-4383-2969}}%
\affiliation{%
 Centro Brasileiro de Pesquisas F\'isicas, Rua Dr. Xavier Sigaud 150, \\
 22290-180 Rio de Janeiro, RJ, Brazil
}%
\affiliation{
 Centro Federal de Educa\c{c}\~{a}o Tecnol\'{o}gica Celso Suckow da Fonseca,  Rodovia M\'{a}rcio Covas, lote J2, quadra J - Itagua\'{i} (Brazil)}

\author{Tiago Castro\orcidlink{0000-0002-6292-3228}}
\affiliation{INAF – Osservatorio Astronomico di Trieste, via Tiepolo 11, I-34131 Trieste, Italy
}%
\affiliation{
 INFN – Sezione di Trieste, I-34100 Trieste, Italy
}%
\affiliation{
 IFPU -- Institute for Fundamental Physics of the Universe, via Beirut 2, 34151, Trieste, Italy
}%
\affiliation{
 ICSC - Centro Nazionale di Ricerca in High Performance Computing,
Big Data e Quantum Computing, Via Magnanelli 2, Bologna, Italy
}%

\date{\today}

\begin{abstract}
The advent of multimessenger cosmology, marked by the detection of GW170817, demonstrated that standard sirens are a valuable cosmological probe. In the absence of an electromagnetic counterpart identification, gravitational waves (GWs) carry valuable information through the dark siren (DS) approach, where the source redshift is estimated using galaxy catalogs of potential hosts within the localisation volume. However, the DS analysis can be affected by galaxy catalog incompleteness at the limits of gravitational-wave detectability, potentially introducing biases in the constraints on cosmological parameters. Focusing on GWs from binary black holes (BBH) detected by the LIGO, Virgo, and KAGRA (LVK) collaboration, we explore the possible systematic biases in the measurement of the Hubble constant ($H_0$). These biases may arise from (1) the incompleteness of catalogs due to the apparent magnitude thresholds of optical telescope sensitivity, and (2) the use of incorrect weighting schemes (e.g., using star formation or stellar mass as tracers of the host galaxy) for each potential host. We found that an unbiased estimate of $H_0$ can be obtained when the corrected weighting scheme is applied to a complete or volume-limited catalog. We use a complete galaxy catalog covering 90\% of the localisation probability ($A_{90\%}$) for each GW detection, employing stellar mass as a tracer. Our results show that a sample of 100 binary black hole events with $A_{90\%}<10\,\rm{deg}^2$ and measured luminosity distances below 1600(2500) Mpc, detected by the LVK at O4(O5) sensitivity, can provide a percent-level measurement of $H_0$, with a precision of approximately 3\%(1\%). This number of detections is expected to be accumulated after approximately 8 and 3 years of observations with the LVK at O4 and O5 sensitivity, respectively. The O5 run provides a reduction in the $H_0$ uncertainty by 1.34 km/s/Mpc compared to the O4-like configuration. The $H_0$ precision increases to approximately $6\%$ when it is assumed that every galaxy has an equal probability of being the host.
\end{abstract}

\keywords{Gravitational waves; cosmology; dark siren analysis; galaxy catalog}
\maketitle


\section{\label{sec:intro}Introduction}
The detection of the first gravitational wave, GW150914 \cite{GW150914}, opened the field of gravitational-wave astronomy. Since that milestone, gravitational waves have emerged as a powerful tool for investigating the expansion history of the Universe and offer insights into the mass and spin distributions of gravitational wave sources \cite{gwtc3_population}. The expansion history can be analyzed using gravitational waves as \textit{standard sirens} initially proposed by Schutz in \citep{schutz}. In the standard sirens context, the GWs alone offer a direct measurement of the luminosity distance. However, the redshift cannot be directly obtained from the GWs, as it is degenerate with the source masses and must therefore be inferred from an associated electromagnetic counterpart. Until now, the only bright standard siren detected is the GW170817 \cite{ligobns}, where the gravitational waves emerge from a merger of two neutron stars at a distance of $40^{+8}_{-14}$ Mpc. Approximately two seconds after the merger, a short gamma-ray burst was detected by two distinct space telescopes: the Fermi Gamma-ray Burst Monitor \cite{Goldstein_2017} and the International Gamma-ray Astrophysics Laboratory \cite{Savchenko_2017}. Additionally, an optical kilonova (e.g. \citet{Arcavi_2017, Coulter_2017, Cowperthwaite_2017, Soares_Santos_2017, Chornock_2017, Kasliwal_2017, Nicholl_2017, Evans_2017, Pian_2017, Smartt_2017, Tanvir_2017, Valenti2017} ) was identified within the region of galaxy NGC 4993, confirming it as the host galaxy. This opens up the possibility of exploring cosmology through gravitational waves, providing a new independent measurement of the local expansion rate of the Universe, known as the Hubble constant, with an uncertainty of 15\% \cite{2017Natur.551...85A}. 

Despite the success of GW170817 to probe cosmology, the absence of other standard sirens, combined with the low number of GW detections from binary neutron stars mergers \cite{gwtc3_published}, discourages their potential to dominate the field of gravitational-wave cosmology. As an alternative, cosmological information can be explored using the ``dark siren" technique. As the name suggests, in this method the electromagnetic counterpart is assumed to be undetectable, and the source redshift is inferred from a statistical technique. The DS approach is further categorized into several distinct methods: (1) the spectral mass that relies on mapping the observed mass spectrum, the redshifted mass, i.e. $m_{z} = (1+z)m$, in luminosity distance to infer the redshift source \cite{Ezquiaga2021, Ezquiaga2022, Farah2025, Magana2024, Borghi2024}, (2) the cross-correlation ones that explored the correlation between the GWs and galaxies distributions to estimate the host galaxy \cite{Mukherjee2021a, Mukherjee2021b, Mukherjee2024, Ghosh2025, Ferri2024}, and (3) the use of all flares in AGNs that might be associated to a BBH merger \cite{bom24bbh, 2021ApJ...914L..34P}. Apart from this, a promising way to identify the host galaxy is to focus on the galaxy catalog \cite{schutz, delpozzo, chen17, fishbach}, where all the galaxies within the GW localisation volume are identified as potential hosts. A probabilistic weight is assigned to each potential host based on its position, which is statistically marginalized over. This approach, known as the \textit{statistical} method, has already been fully explored with real events of gravitational waves \cite{fishbach, darksiren1, palmese20_sts, palmese2023, Ballard2023, alfradique2024,bom2024}, the latest measurement\cite{bom2024} showed that from a catalog with 15 dark sirens it is possible to find a measure of $H_0$ with precision of 18\% and when combined with the bright siren GW170817 is reduced to 6\%.

The expectation is that the next generation of interferometers leads to a $H_0$ precision of $<1\%$ \cite{Borhanian2020, Muttoni2023, Zhu2023}, which will require more rigorous control over potential systematic effects in DS cosmology. The study by \cite{Gair2023} provides an extensive analysis, demonstrating that the statistical DS method, when applied to consistent simulated data, yields cosmological measurements without bias. In a mock data analysis, \cite{Gray2020} also showed that the method is capable of providing unbiased inference for $H_0$. The study further explored the effects of galaxy incompleteness and the weighting of host galaxies on cosmological inferences. Recently studies \cite{Hanselman2025, Perna2024} investigated the influence on $H_0$ when different galaxy properties are used as hosting probability, both works showed that a systematic bias is introduced into the $H_0$ estimation if the host probability is mismatched. Another potential source of systematic errors in the DS analysis is the mismodeling of the galaxy redshift uncertainty. In their study, \cite{Turski2023} addressed this issue by modeling the redshift uncertainty using a Gaussian and a modified Lorentzian function. They showed that the systematic errors introduced by this approach are much smaller than the statistical ones. On the other hand, the probability distribution function of the photometric redshift for galaxy catalogs from wide-field surveys can be estimated using deep learning techniques \cite{Lima2021, teixeira2024, Li2024}. The resulting distribution may exhibit more complexity than a simple Gaussian or Lorentzian form, highlighting the need to explore these systems beyond traditional models.

In this work, we investigate the systematic biases in the measurement of $H_0$ that may arise from statistical DS analyses. We focus on the effects of magnitude-limited catalogs and incorrect weighting of GW host galaxy probabilities, particularly those based on stellar mass (SM) and star formation rate (SFR). Although previous studies have explored these issues, our analysis extends the investigation in several key aspects. We used simulated sky maps from \texttt{BAYESTAR} \cite{bayestar}, allowing a more realistic representation of measurements derived from gravitational wave detections. Furthermore, we go beyond traditional magnitude-limited samples by incorporating a volume-limited sample and investigating the bias that could arise from galaxy catalog incompleteness. In contrast to the analytical approximations often used in prior works, we quantify the selection function through a simulation-based approach, enhancing the robustness of our results. Finally, we analyze the impact on the precision of $H_0$ due to uncertainties in the GW luminosity distance, the 90\% credible sky area, and the GW redshift range. This paper is organized as follows: In Section \ref{method}, we present the statistical DS methodology. In Section \ref{mocksec}, we describe the simulation prescription for gravitational wave detections with their corresponding sky maps, and the mock galaxy catalog used. Our results are presented in Section \ref{results}, and we summarize and conclude in Section \ref{conclusion}.

\section{Dark siren bayesian analysis}\label{method}
Using the Bayes' theorem, the $H_0$ posterior for a given \textit{i}-th GW event, with observed data $d_{\rm GW}^{i}$, can be written as \cite{Gair2023}:
\begin{eqnarray}
  p(H_0|d_{\rm GW}^{i}) &\propto& p\left(H_0\right)\mathcal{L}\left(d_{\rm GW}^{i}|H_0\right), \\
  \label{likedef}
    \mathcal{L}\left(d_{\rm GW}^{i}|H_0\right) &\propto& \frac{\int dz \mathcal{L}_{\rm{GW}}\left(\hat{d}^{i}_{L}|d_{L}\left(z,H_0\right)\right)p_{\rm{CBC}}\left(z\right)}{\int dz P^{\rm{GW}}_{\rm{det}}\left(z,H_0\right)p_{\rm{CBC}}\left(z\right)},
\end{eqnarray}
where $P^{\rm{GW}}_{\rm{det}}$ is the GW detection probability and it describes the probability of a GW source located at a redshift \textit{z} to be detected based on specific detection conditions, $\mathcal{L}_{\rm{GW}}$ is the GW likelihood of measuring a luminosity distance $\hat{d}_{L}^{i}$ given a $H_0$ value and a galaxy at a redshift \textit{z}, and $p\left(H_0\right)$ is the prior distribution. $p_{\rm{CBC}}$ is the compact binary coalescence probability.

\begin{figure}
    \includegraphics[width=\linewidth]{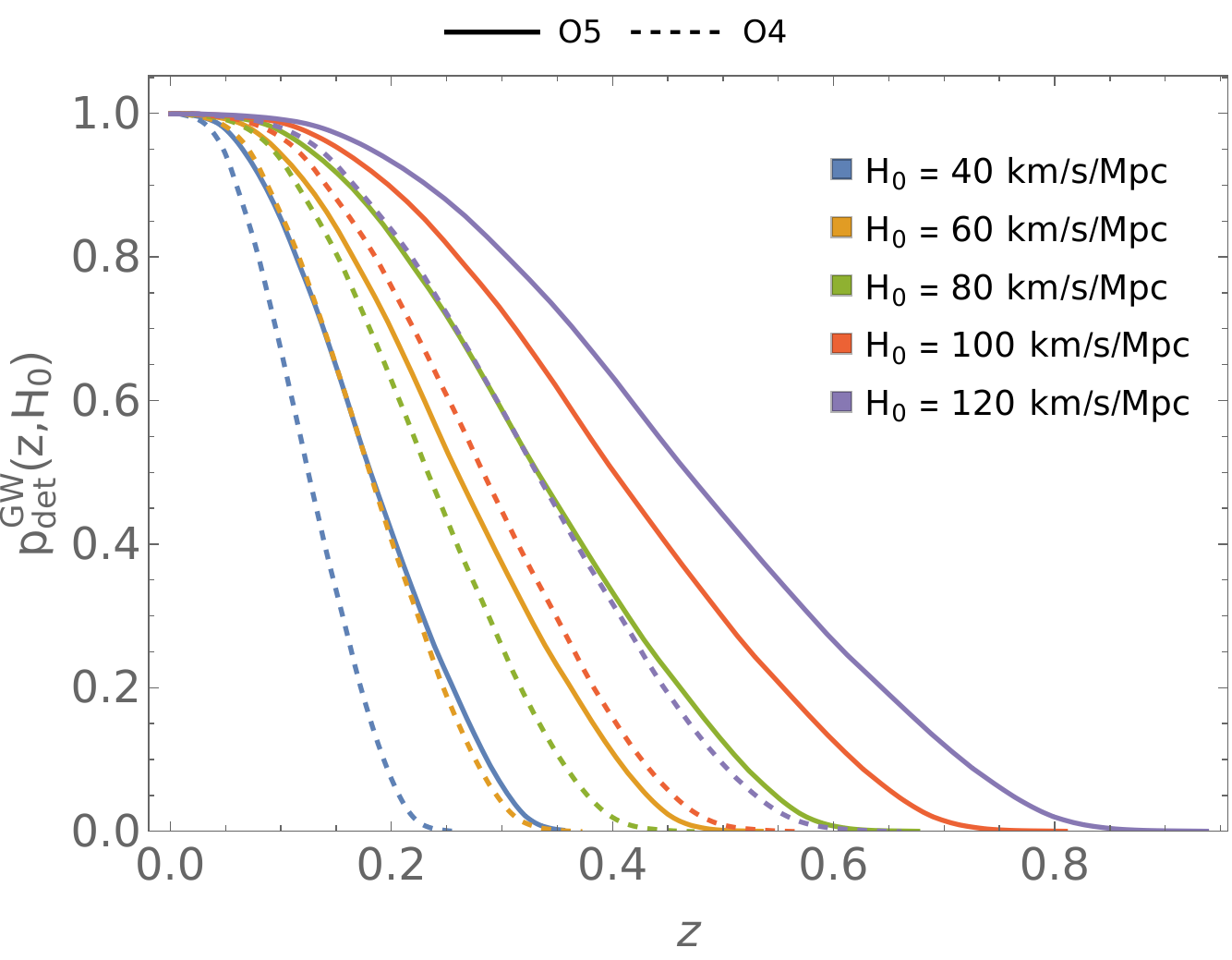}
    \caption{GW detection probability as a function of \textit{z}, with each color representing a different value of $H_0$. These results were generated using the design sensitivity curves for O4 (dashed lines) and O5 (solid lines), a BBH population, and assuming that the detection criteria are a network SNR above 12, a 90\% confidence localisation area ($A_{90\%}$) smaller than 100 deg$^2$, and measured luminosity distance below 1600 Mpc for O4 (2500 Mpc for O5).}
    \label{fig:selfunc}
\end{figure}

As defined in \cite{Gair2023}, the detection probability $P_{\rm det}^{\rm GW}\left(z, H_0\right)$ accounts for the fact that the analysis is limited to GW events detected above a certain threshold, $d_{\rm thr}$. It is expressed as the integral of the GW likelihood over the data $d_{\rm GW}$, conditioned on $d_{\rm GW}$ exceeding the detection threshold $d_{\rm thr}$. 
The calculation of the term $P_{\rm det}^{\rm GW}$ requires integrating over all the realisations of GW events based on the binary population model parameters, considering different values of \textit{z} and $H_0$, and applying the detection condition. To achieve this, we follow the approach outlined in \cite{Gray2020} and employ a Monte Carlo integration defined as
\begin{eqnarray}
     P_{\rm det}^{\rm GW}\left(z, H_0\right) &=& \frac{1}{N_{\rm GW}}\sum_{i=1}^{N_{\rm GW}}P_{\rm det}^{\rm GW}\left(d_{\rm GW}^{i}, z, H_0\right),
\end{eqnarray}
where $d_{\rm GW}$ represents the GW data corresponding to an event (whether detected or not), which is associated with source parameters (such as inclination, right ascension, declination, component masses, and polarization) that may influence the detectability of the event. In the equation above, we have excluded the dependence on the sky position, assuming that the probability is approximately uniform across the sky. The term $P_{\rm det}^{\rm GW}\left(d_{\rm GW}^{i}, z, H_0\right)$ is assumed to be equal to one if the GW event $d_{\rm GW}^{i}$ satisfies the detection condition, and zero otherwise. The detection criterion is that the network signal-to-noise ratio (SNR) must be greater than 12 and 45 for O4 and O5 runs, respectively; the measured luminosity distance must be below 1600(2500) Mpc for the O4(O5) observing run; and an additional condition is applied to the sky localisation area to ensure the selection of well-localized events. For this, we consider two configurations: $A_{90\%}<10\,\rm{deg}^{2}$ or $A_{90\%}<100\,\rm{deg}^{2}$. Thus, $P_{\rm det}^{\rm GW}\left(z, H_0\right)$ is a smooth function whose values range from 1 to 0. The Monte Carlo equation, as defined above, was computed using simulations of 100,000 GW events from BBH mergers, following the procedure and assumptions for the population properties described in subsection \ref{gwsimu}. These GW simulations were conducted for 17 distinct values of $H_0$ within the prior range of [40, 120] km/s/Mpc. The resulting function was then interpolated to ensure it was continuously defined across the entire prior range. We computed the $P_{\rm det}^{\rm GW}\left(z, H_0\right)$ function for each adopted detection condition configuration. Figure \ref{fig:selfunc} presents $P_{\rm det}^{\rm GW}\left(z, H_0\right)$ as a function of \textit{z}, computed for different values of $H_0$ values within the redshift range of interest for this analysis, assuming O4 and O5 scenarios. 

The term $p_{\rm{CBC}}$ can be modelled as described in \citet{Gray2023, Mastrogiovanni2023, Borghi2024, Perna2024},  
\begin{equation}
    \label{pcbc}
    p_{\rm CBC}\left(z\right) = \frac{\int dx_{\rm gal} p_{\rm rate}\left(\textit{z}, x_{\rm gal}\right)p_{\rm cat}\left(\textit{z}, x_{\rm gal}\right)}{\int d\textit{z}dx_{\rm gal} p_{\rm rate}\left(\textit{z}, x_{\rm gal}\right)p_{\rm cat}\left(\textit{z}, x_{\rm gal}\right)}
\end{equation} 
with $p_{\rm rate}$ represent the probability that a galaxy with redshift \textit{z} and specific properties $x_{\rm gal}$ hosts a GW event and $p_{\rm cat}$ is the probability of a galaxy with these properties being located at redshift \textit{z}, essentially corresponding to the galaxy catalog probability. Here, we model the GW host probability $p_{\rm rate}$ based on the galaxy properties, specifically the stellar mass $M_{*}$ or the star formation rate $\rho_{\rm SFR}$, i.e. $x_{\rm gal}=\{M_{*}, \rho_{\rm SFR}\}$. This probability is defined in the detector frame; therefore, it must be multiplied by $\left(1 + z\right)^{-1}$ to account for the time dilation caused by the expansion of the Universe between the source and the detector frames. We assume that the galaxy catalog is complete and provides the precise values of $\{M_{*}, \rho_{\rm SFR}, z\}$, such that $p_{\rm cat}$ can be written as a delta function
\begin{eqnarray}
    \label{pcat}
    p_{\rm cat}\left(z, x_{\rm gal}\right) &\approx& \frac{1}{N_{\rm gal}}\sum_{i=1}^{N_{\rm gal}}\delta\left(z-\hat{z}^{i}_{\rm gal}\right)\delta\left(x_{\rm gal}-\hat{x}_{\rm gal}^{i}\right),
\end{eqnarray}
where $N_{\rm gal}$ is the number of galaxies in the catalog. The marginal GW likelihood was computed assuming a Gaussian function as described in \cite{bayestar}:
\begin{eqnarray}\label{gaussianlike}
\mathcal{L}_{\rm{GW}}\left(\hat{d}^{j}_{L}|d_{L}\left(z,H_0\right),\hat{\Omega}_{i}\right)\approx\frac{p(\hat{\Omega}_{i})}{\sqrt{2\pi}\sigma(\hat{\Omega}_{i})}\nonumber\\
\times\exp\left[-\frac{(d_{L}\left(z,H_0\right)-\hat{d}_{L}^{j}(\hat{\Omega}_{i}))^2}{2\sigma^{2}(\hat{\Omega}_{i})}\right],
\end{eqnarray}
where the $p\left(\hat{\Omega}_{i}\right)$ is the position probability, while $\hat{d}_{L}^{j}\left(\hat{\Omega}_{i}\right)$ and $\sigma\left(\hat{\Omega}_{i}\right)$ denote the mean and standard deviation of the luminosity distance, respectively. 
The solid angle of the \textit{i}-th HEALPIX pixel is given by $\hat{\Omega}_{i}$.

By substituting eq.\ref{pcat} into eq.\ref{pcbc}, we obtain the final form of $p_{\rm{CBC}}\left(z\right)$ as
\begin{eqnarray}
    p_{\rm CBC}\left(z\right) \propto \frac{1}{N_{\rm gal}} \sum_{i=1}^{N_{\rm gal}} p_{\rm rate}\left(z, \hat{x}_{\rm gal}^{i}\right)\delta\left(z-\hat{z}^{i}_{\rm gal}\right),
\end{eqnarray}
that simplifies the likelihood (eq.\ref{likedef}) for the measured luminosity distance corresponding to the \textit{j}-th gravitational wave event to:
\begin{eqnarray} \mathcal{L}\left(d^{j}_{\rm GW}|H_0\right)\propto     \sum_{i=1}^{N_{\rm gal}}\mathcal{L}_{\rm{GW}}\left(\hat{d}^{j}_{L}|d_{L}\left(\hat{z}^{i}_{\rm gal},H_0\right),\hat{\Omega}_{i}\right)\nonumber\\ 
\times p_{\rm{rate}}\left(\hat{z}^{i}_{\rm gal}, \hat{x}_{\rm gal}^{i}\right)/
     \sum_{i=1}^{N_{\rm gal}}P^{\rm GW}_{\rm det}\left(\hat{z}^{i}_{\rm gal}, H_{0}\right)p_{\rm{rate}}\left(\hat{z}^{i}_{\rm gal}, \hat{x}_{\rm gal}^{i}\right),\hspace{0.4cm}
\end{eqnarray}
where $\hat{\Omega}_{i}$ represents the solid angle of the HEALPIX pixel in which the \textit{i}-th galaxy is located.

For a set of $N_{\rm GW}$ luminosity distances observed from gravitational waves, $\{\hat{d}_{L}\}\left(\equiv\{d_{\rm GW}\}\right)$, and assuming the independence of each data, the above equation is extended to:
\begin{equation} 
\mathcal{L}\left(\{\hat{d}_{L}\}|H_0\right) \propto \prod_{j=1}^{N_{\rm GW}} \mathcal{L}\left(\hat{d}^{j}_{L}|H_0\right).
\end{equation}

\section{Mock data}\label{mocksec}

\subsection{Mock galaxy catalog}

This study employs the \texttt{\small Magneticum} galaxy catalog \cite{Hirschmann2014, Ragagnin2017} as our mock galaxy dataset. The \texttt{\small Magneticum} simulation suite contains several cosmological hydrodynamic simulations that account for the evolution of dark and luminous matter under the influence of galaxy and baryonic processes. The galaxy catalogs used in this work were extracted from the boxes 2 and 2b \footnote{\url{http://www.magneticum.org/simulations.html}}. These simulations are performed within a comoving periodic volume of (500 Mpc)$^3$ for box 2 and a 6 times larger volume for box 2b. The lightcone covers an area of 5143 deg$^2$ (a 1/8th of a sky-wedge) and extends to a redshift of \textit{z}=1.3. The simulations assumes a standard $\Lambda$CDM cosmological model, with the following parameters: $\Omega_{m} = 0.272$, $\Omega_{b} = 0.0456$, $\Omega_{\Lambda} = 0.728$, $n=0.963$, $H_0 = 70.4$ km/s/Mpc, and $\sigma_8 = 0.809$. The lightcone was generated by extracting galaxies from snapshots of the simulation at various redshifts, replicating the box when needed to ensure a consistent representation of the cosmic volume across the observational limits. The galaxy properties, including stellar mass, star formation rate, metallicity, and absolute magnitude, were derived directly from the simulation output. These quantities were computed based on detailed models of gas dynamics, stellar evolution, and feedback processes, providing a comprehensive view of the galaxy population at different epochs. The \texttt{\small Magneticum} catalog offers a diverse range of galaxy properties, which are crucial for our analysis to consistently describe the large-scale structure, galaxy evolution, and cosmological parameter estimation.

\subsection{GW simulations}\label{gwsimu}

In order to compute the $H_0$ posterior under different observational conditions, the Bayesian framework described in Section \ref{method} was applied to a catalog of simulated BBH gravitational wave detections, along with galaxy information from the \texttt{\small Magneticum} catalog. The gravitational wave detections were simulated using the Python package \texttt{\small ligo.skymap}\footnote{\url{https://lscsoft.docs.ligo.org/ligo.skymap/}{https://lscsoft.docs.ligo.org/ligo.skymap/}}, which includes the \texttt{\small BAYESTAR} code, an efficient tool for reconstructing gravitational wave sky localisation. We perform an end-to-end simulation of 100,000 BBH mergers for the LIGO and Virgo network detector with sensitivity comparable to that attained in the O4 and O5 runs. The amplitude spectral densities used for the O4 run are available at \url{https://git.ligo.org/lscsoft/gwistat/-/tree/master/psd?ref_type=heads}. For the O5 run, we used the public sensitivity curves presented in \cite{2018LRR....21....3A} and available at \url{https://dcc.ligo.org/LIGO-T2000012/public}. For the LIGO detectors, we used the curve labeled as ``AplusDesign"; for Virgo, ``avirgo\_O5low\_NEW"; and for KAGRA, ``kagra\_80Mpc". The BBH population follows a mass distribution defined by the power law plus peak model with parameters fixed to the best values found in \cite{gwtc3_population}.

To simulate an injected GW event, a galaxy from the \texttt{\small Magneticum} catalog was first selected as the host, with the GW host probability $p_{\rm rate}$ weighted by stellar mass, star formation rate or equally, i.e. $p_{\rm rate} \propto \{M_{\star}, \rho_{\rm SFR}, 1\}$. In this way, the 3D position (right ascension, declination, and redshift) of each injected GW event was set to match the true sky position of its host galaxy. We assume the IMRPhenomD frequency-domain waveform model to produce the GW signals using the \texttt{\small BAYESTAR} software. After generating the injections, a matched-filter analysis was performed to compute the SNR for each injected event. An event was considered detected if the network SNR is above 12 and at least 2 detectors have a single–detector SNR above 4. Finally, the 3-dimensional sky map was reconstructed for all the detections GW event with the \texttt{\small BAYESTAR}. To select the best events for the DS analysis in each observing run, we adopted the following criteria: (1) for the O4 run, GW events with an measured luminosity distance $d_{\rm{L, mea}}<1600\,\rm{Mpc}$ and an SNR above 12, and (2) for the O5 run, GW events with an measured luminosity distance $d_{\rm{L, mea}}<2500\,\rm{Mpc}$ and an SNR above 45. The choice to use the measured, rather than the true, luminosity distance ensures consistency with the hierarchical Bayesian likelihood and mitigates potential biases in the analysis, as discussed in \cite{Essick2024}. In this way, despite the O5 configuration selecting more distant GW events, possibly with larger uncertainties in luminosity distance, the more restrictive SNR cut ensures that the distribution of luminosity distance uncertainties is equivalent in both observing run configurations. The GW events used in the analysis were also selected based on their 90\% confidence interval (CI) sky localisation area. The detection configurations were divided into: (1) events with $A_{90\%}<100\,\rm{deg}^2$, and a more restrictive set, (2) events with $A_{90\%}<10\,\rm{deg}^2$.

\begin{figure*}
    \centering
    \begin{tabular}{m{0.5\textwidth} m{0.5\textwidth}}
        \parbox[c]{\linewidth}{%
            \includegraphics[width=\linewidth]{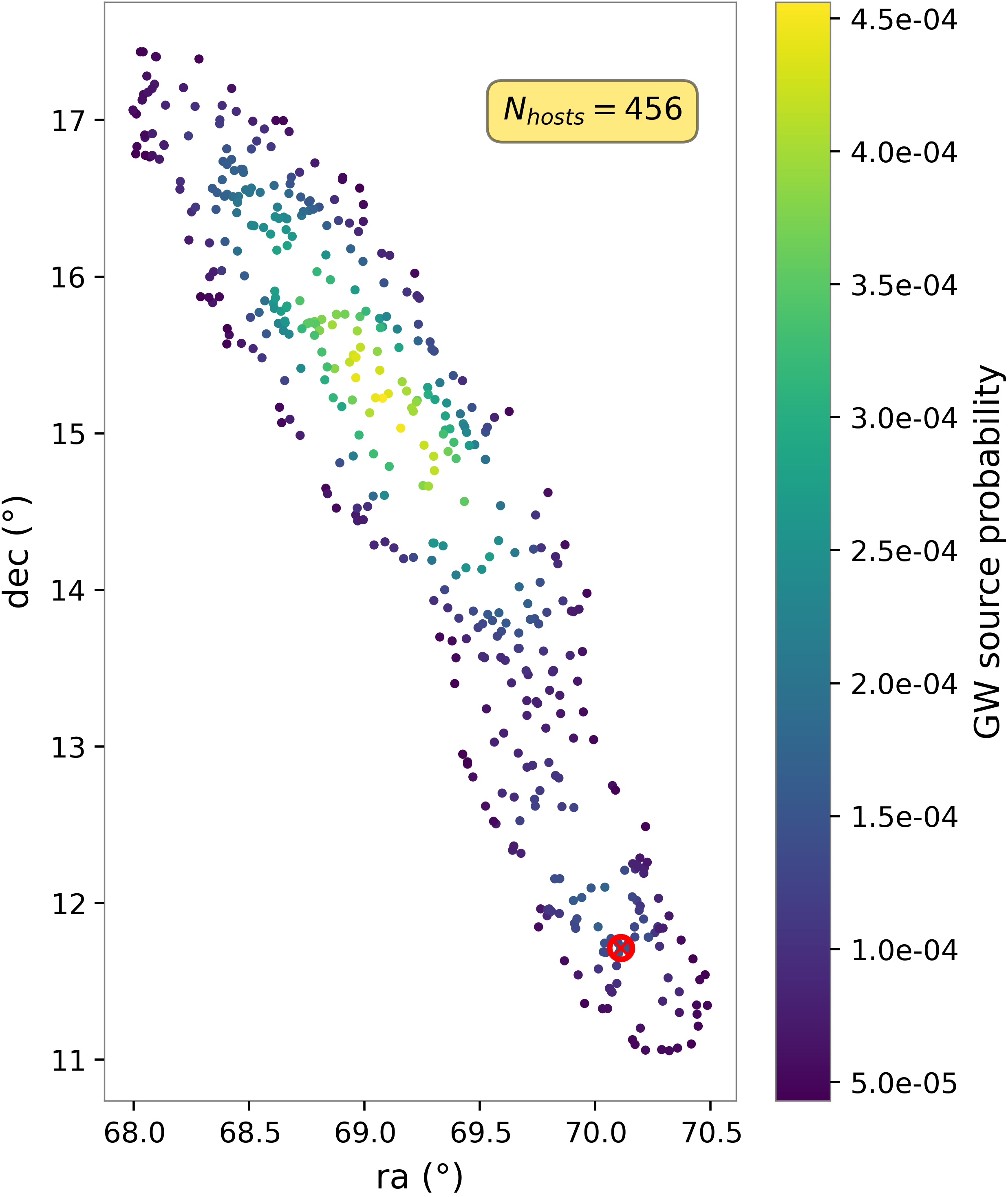}
        }
        & 
        \parbox[c]{\linewidth}{
            \includegraphics[width=.8\linewidth]{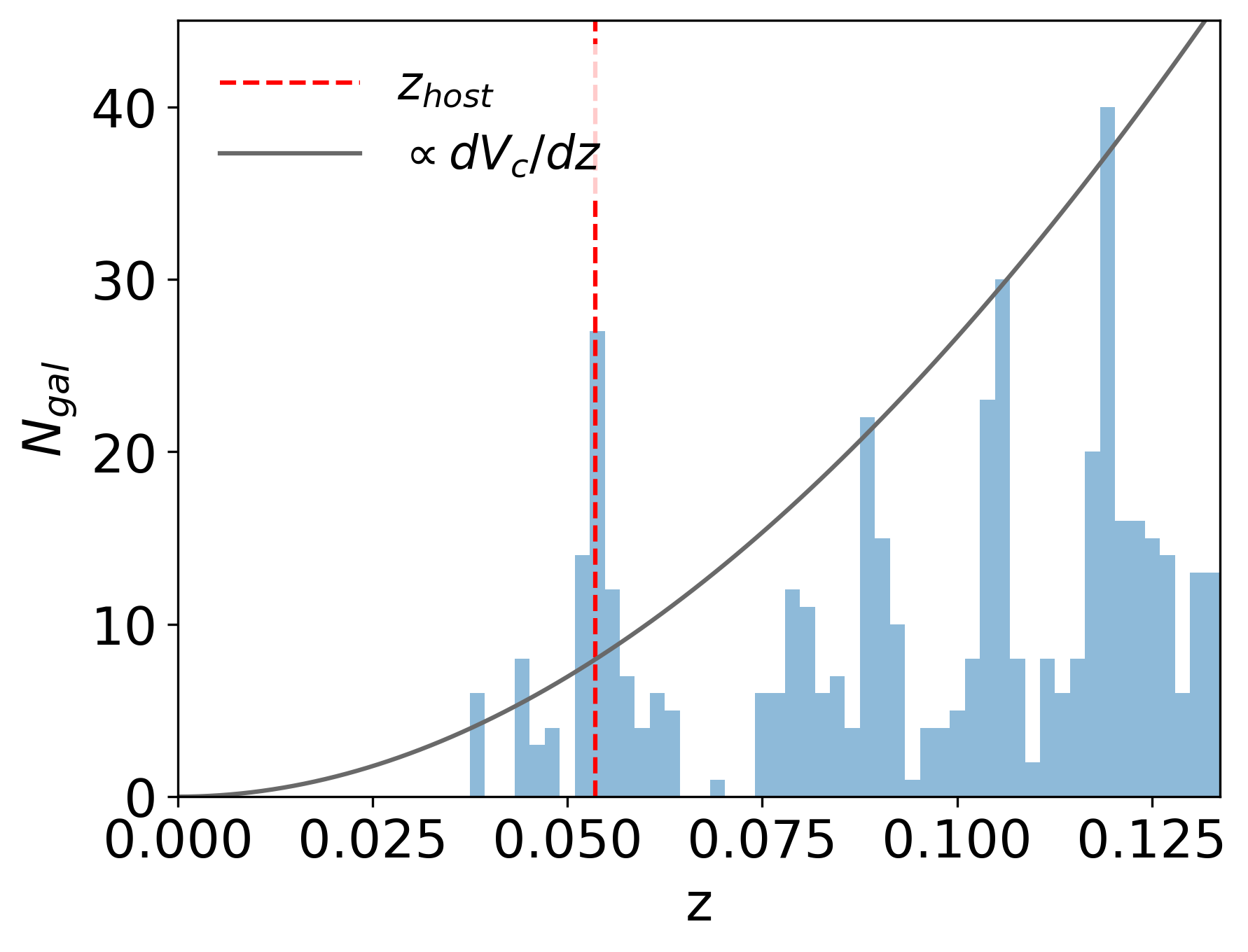} \\[0.5em] 
            \includegraphics[width=.8\linewidth]{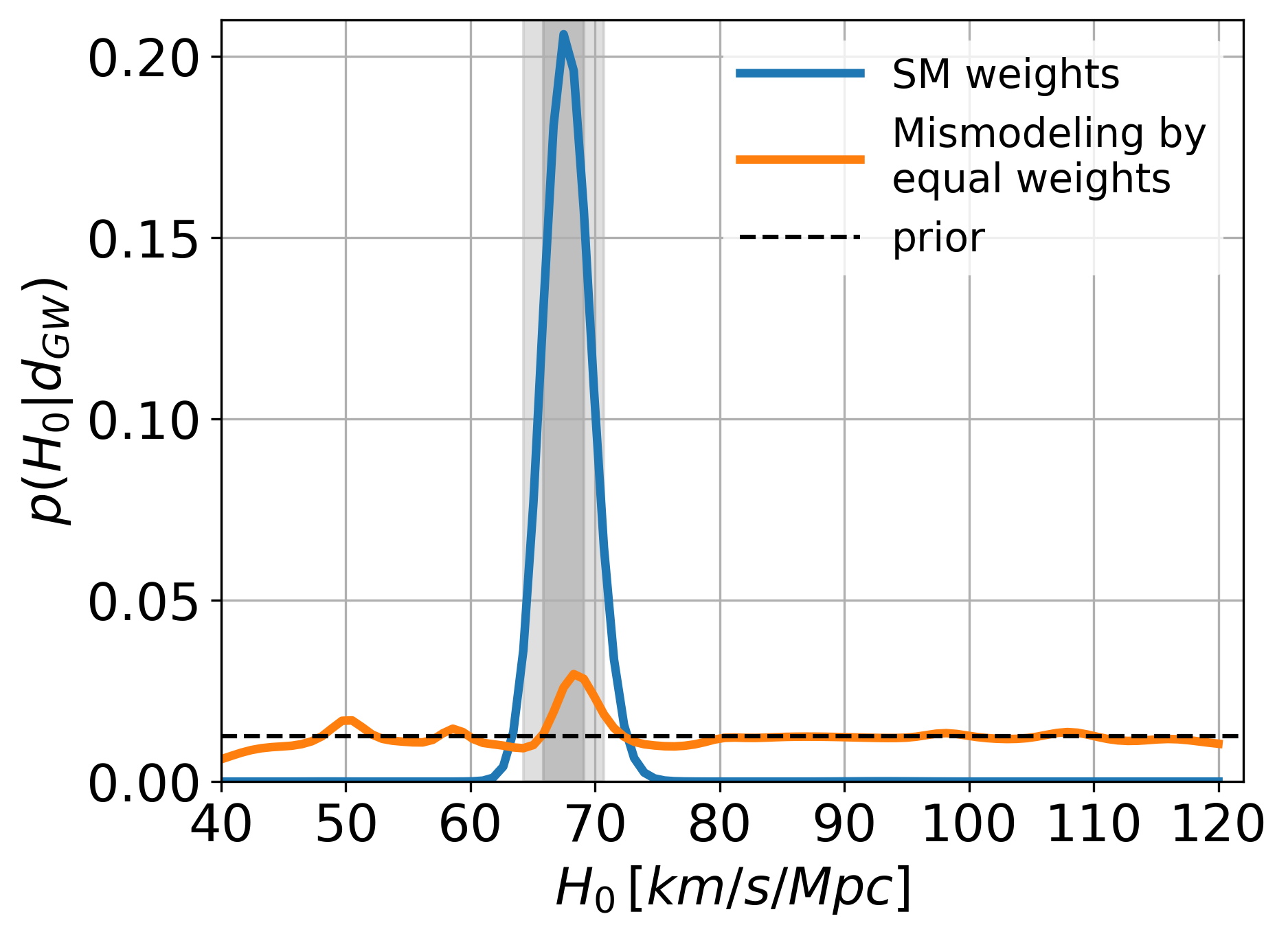}
        } \\
    \end{tabular}
\caption{\textit{Left}: The sky localisation of the potential host galaxies of a GW example case from our mock catalog. Each point represents a galaxy, colored by the positional probability of the GW source, computed with \texttt{BAYESTAR}. The red circle with a cross indicates the true position of the host galaxy.
\textit{Right}: The redshift distribution of the potential host galaxies, extracted from \texttt{\small Magneticum} for this particular GW event, is shown in the top panel. The dashed red line represents the $z$ value of the true host, and the gray solid line represents a uniform distribution in comoving volume. The bottom panel shows the $H_0$ posterior for the DS method, assuming, incorrectly, equal weights (in orange) and weighting the galaxies by their stellar mass (in blue).}
\label{fig:event_example}
\end{figure*}

After constructing the three-dimensional sky map for the gravitational wave detection sample, we perform a cross-match between the resulting localisations and the \texttt{\small Magneticum} galaxy catalog. Although the sky-localisations provide the luminosity distance posterior along each line-of-sight, most wide survey catalogs that probe the universe at extragalactic distances lack precise distance measurements, primarily due to their dependence on an intrinsic calibration process (as seen in the \textit{Standard Candles}). Therefore, to obtain results closer to the observed universe, we constructed a three-dimensional subcatalog containing all galaxies within the 90\% credible area. This subcatalog includes galaxies whose redshift values correspond to the 90\% CI bounds of the luminosity distance, with the minimum and maximum redshifts determined by assuming the lower and upper values of $H_0$ from our prior, respectively. In our 100 GW sub-catalogs, we find that, on average, approximately four GW events have their true host galaxies located outside the 90\% CI. Although these true hosts are not included in the 90\% CI, the posterior obtained using the 90\% CI remains statistically consistent with that obtained utilizing the broader 99\% CI. As an example, the left panel of Fig. \ref{fig:event_example} presents all the galaxies within the 90\% CI of the sky localisation area for a specific simulated GW event. The top panel of Fig. \ref{fig:event_example} shows the redshift distribution of these potential host galaxies, with the true redshift of the host galaxy indicated by a dashed line. Finally, we applied this subcatalog to each simulated GW detection using the statistical method outlined in Section \ref{method} to estimate the Hubble constant. 
To explore the effect of galaxy incompleteness on the $H_0$ measurement, no corrections for completeness were applied in the calculations. The $H_0$ posterior for the example GW event illustrated in Fig. \ref{fig:event_example} is shown in the bottom panel. In these results, we see that, as the position of the true host galaxy coincides with a region of overdensity, the $H_0$ posterior distribution is consistent with the fiducial value of 70.4 km/s/Mpc.

\section{Result and discussion}\label{results}
In this section, we present our estimation of the Hubble constant for the LVK network during the O4 and O5 runs, using the simulated mock data in the statistical framework described in Section \ref{method}.
We performed an analysis considering four distinct configurations of apparent magnitude completeness, characterized by magnitude cuts of $m < \{22, 21, 20, 19\}$), two detector sensitivities mimicking the O4 and O5 runs, and three different weighting schemes for the host galaxy. These schemes assume: (1) equal probability for all galaxies in the catalog to host the GW source, (2) stellar mass, and (3) star formation as a tracer for the host galaxy. In practice, since stellar mass and star formation are not directly observed, it is common to use a galaxy's K-band and B-band luminosities as proxies for stellar mass and star formation, respectively. For each configuration, we performed $30$ independent realisations of the combined posteriors on $H_0$ using a sample of 100 DSs, analyzing the precision and quantifying the biases relative to the injected $H_0$ value. This approach ensures that our results are not dependent on a single realisation of the population, allows us to assess the statistical fluctuations due to population realisations, and leads to more robust conclusions. We adopt a uniform prior on $H_0$ in the range $\left[40,120\right]$ km/s/Mpc.

In Table \ref{tab:a9010deg}, we present the final constraint on the Hubble constant for DSs selected based on the 90\% CI of the localisation probability being below 10 deg$^2$, except where explicitly stated otherwise. 
We define the bias of the $H_0$ estimator as the mean difference between the peak of the combined posterior, obtained from a sample of 100 GWs events, and the fiducial value from the \texttt{\small Magneticum} simulation ($H_0^{\rm fid}=70.4$ km/s/Mpc), computed over 30 independent realisations. Specifically, let \( \hat{H}_0^{(i)} \) denote the posterior peak from the combined posterior in the \( i \)-th realisation. The bias is then given by:
\begin{equation}
\label{biaseq}
\Delta H_0 = \frac{1}{30} \sum_{i=1}^{30} \left( \hat{H}_0^{(i)} - H_0^{\rm fid} \right).  
\end{equation}
The final $H_0$ measurement was obtained by averaging the posterior distributions from all 30 realisation of a sample of 100 GWs. For the computation of the completeness fraction (see its definition in equation \ref{eqcompleteness}), we used the reference distance corresponding to the detection threshold: O4 $d_{\rm{L,ref}} = 1600$ Mpc for O4 and $d_{\rm{L,ref}} = 2500$ Mpc for O5. 
Our results indicate that is possible to obtain a unbiased final $H_0$ constraint for a galaxy catalogue volume-limited, even under a restrictive apparent magnitude threshold, $m<19$, where many distant galaxies have already been removed. We found that the $H_0$ bias increases with the apparent magnitude threshold adopted for the O4 run, reaching a maximum of $\Delta H_0 = 0.32 \pm 1.78$ km/s/Mpc for $m < 19$. In contrast, for the O5 run, the $H_0$ bias remains constant at approximately 0.22 km/s/Mpc across all apparent magnitude thresholds considered. This demonstrates that the volume-limited catalog is capable of effectively representing the true distribution of galaxy overdensities and underdensities along the line of sight. Furthermore, the results for different galaxy weighting schemes indicate that even if the wrong model is assumed, the final $H_0$ constraint remains unbiased. However, both the uncertainty and bias in the $H_0$ measurement increase in this mismodeling scenario, suggesting that a biased measurement may occur in some cases. This will be explored in detail in the subsection \ref{mismodeling}. The maximum bias magnitude of $H_0$ occurs when it is assumed that all galaxies have an equal probability of hosting the GW event, but the SM is incorrectly applied as the weighting scheme. We found a bias of $3.67\pm 8.68$ km/s/Mpc for O4, which increases to $5.16\pm 16.13$ km/s/Mpc for O5. The minimum bias occurs when the true BBH merger distribution follows SM but is recovered using SFR, resulting in a bias of $0.71\pm 1.90$ km/s/Mpc ($-0.18\pm 0.42$ km/s/Mpc) for O4 (O5). The improvement achieved with SFR, compared to equal weighting, can be attributed to the positive correlation between SM and SFR (see Fig. \ref{fig:cornerplot}), which facilitates the correct assignment of probability for the majority of possible host galaxies. 

\begin{figure*}
    \includegraphics[width=\textwidth]{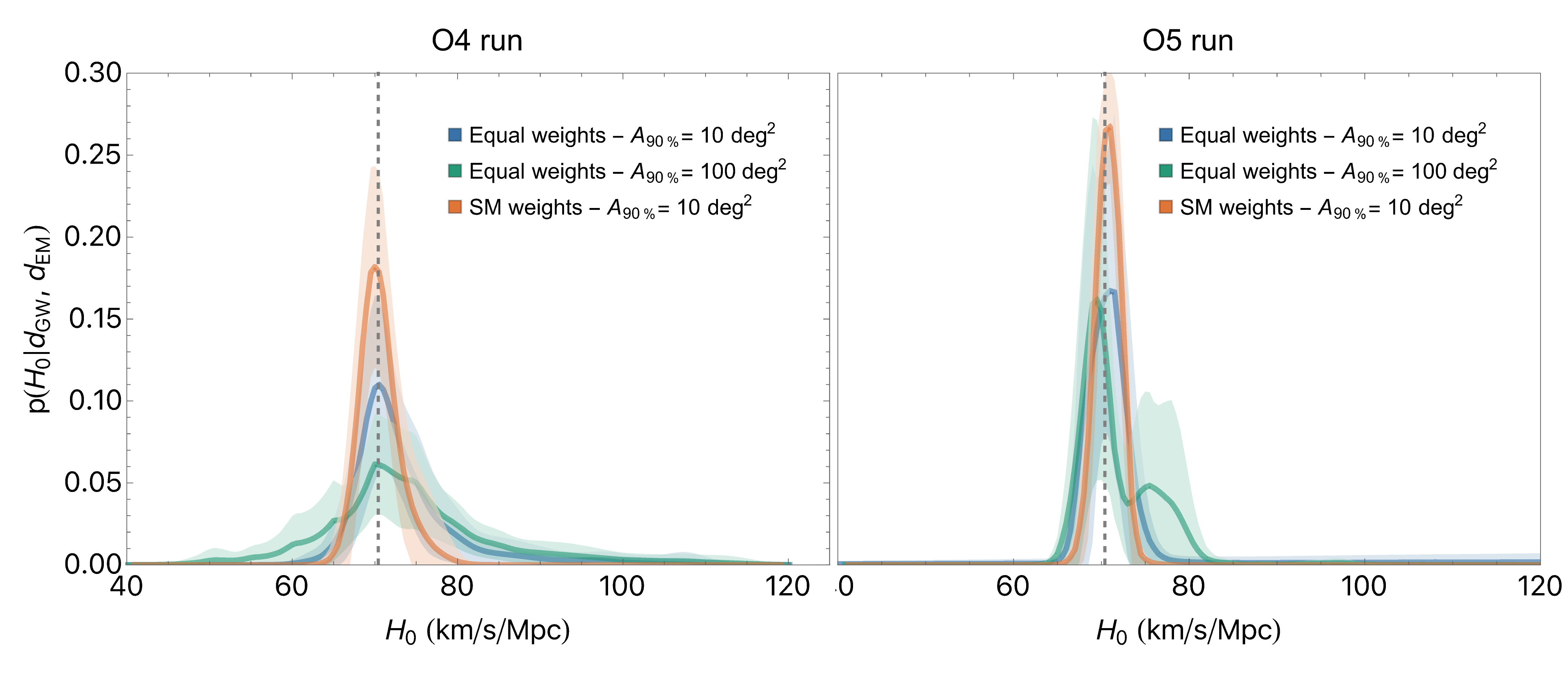}
    \caption{Comparison of $H_0$ posteriors between three different configurations for O4-like (left panel) and O5-like (right panel) configuration: (1) in blue, using the fact that all galaxy in the catalog are equally likely to be the host and selected only the best-localized DS events (localisation area is less than 10 deg$^2$), (2) in green, also assuming the same weigh scheme of (1) but selected only events with $A_{90\%}<100$ deg$^2$, and (3) in orange, using the best-localized DS events weighting by the stellar mass. The shaded area and the thick line represent the 68\% CI and the mean of the 30 realisation. The gray dashed line is the fiducial $H_0$ value from the \texttt{\small Magneticum} simulation.}
    \label{fig:DS04}
\end{figure*}

Figure \ref{fig:DS04} shows the combined $H_0$ posterior for 100 simulated DSs detected in an O4- and O5-like configuration. The addition of the SM weighting reduces the uncertainty in the $H_0$ measurement (compare the green and blue curves), an improvement of a factor of 1.97 for O4 and 1.89 for O5 run. All the estimations are in agreement with the fiducial $H_0$ value. In the best-case scenario, assuming stellar mass weighting, $H_0$ can be constrained to 3\% (1\%) for O4-like (O5-like) configurations. The comparison between the O4 and O5 results shows that the next terrestrial detector sensitivity can cause a reduction of 2.52 km/s/Mpc and 1.34 km/s/Mpc in the $H_0$ uncertainty for the equal and SM weighting schemes, respectively. This shows that the benefit of enhanced detector sensitivity is comparable to the improvements achieved by using SM as the GW host tracer. 

\begin{table*}
\caption{$H_0$ measurements for differents GWs detections configurations (a weighting host galaxies by the stellar mass or equal weighting, different apparent magnitude thresholds, and two cuts over the measured luminosity distance, $d_{\rm{L,mea}}<\{1600, 2500\}\, \rm{Mpc}$) across $30$ realisations of a sample of 100 GWs with $A_{90\%}<10\, \rm{deg}^2$, except where explicitly stated otherwise. The galaxy completess is defined as written in equation \ref{eqcompleteness}. The Hubble
parameter $H_0$, its uncertainties, and the bias ($\Delta H_0$) are given in units of km/s/Mpc. The $H_0$ bias is defined in equation \ref{biaseq}.}
\begin{ruledtabular}
\begin{tabular}{ccccccccc}
 & \multirow{2}{*}{mth} & Skymap & \multirow{2}{*}{Completeness} & \multirow{2}{*}{$H_0$} & \multirow{2}{*}{$\sigma_{H_{0}}$ } & \multirow{2}{*}{\large{$\frac{\sigma_{H_{0}}}{H_{0}}$}} & \multirow{2}{*}{$\Delta H_0$} & \multirow{2}{*}{$\frac{\Delta H_0}{\sigma_{H_{0}}}$}\\
 & & coverage & & & & &\\ 
 \hline\hline
 $\mathbf{d_{\rm{L,mea}} < 1600\, \rm{Mpc}}$ and SNR$>$12 - O4\\
\hline 
No weighting & - & 100\% & 100\% & $70.60^{+5.56}_{-3.41}$ & 4.49 & 0.06 & 1.84$\pm$3.43 & 0.41$\pm$0.76\\
No weighting ($A_{90\%}<100\, \rm{deg}^2$) & \multirow{2}{*}{-} & \multirow{2}{*}{100\%} & \multirow{2}{*}{100\%} & \multirow{2}{*}{$71.00^{+10.40}_{-7.72}$} & \multirow{2}{*}{9.06} & \multirow{2}{*}{0.13} & \multirow{2}{*}{0.68$\pm$5.78} & \multirow{2}{*}{0.08$\pm$0.64}\\
No weighting & \multirow{2}{*}{22.0} & \multirow{2}{*}{99.9\%} & \multirow{2}{*}{98\%} & \multirow{2}{*}{$70.00^{+4.42}_{- 2.86}$} & \multirow{2}{*}{3.64} & \multirow{2}{*}{0.05} & \multirow{2}{*}{0.98$\pm$1.94} & \multirow{2}{*}{0.27$\pm$0.53}\\
(magnitude-limited) & & & & & & & \\
No weighting & \multirow{2}{*}{21.0} & \multirow{2}{*}{99.4\%} & \multirow{2}{*}{86\%} & \multirow{2}{*}{$70.00^{+4.07}_{-3.16}$} & \multirow{2}{*}{3.61} & \multirow{2}{*}{0.05} & \multirow{2}{*}{0.55$\pm$2.07} & \multirow{2}{*}{0.15$\pm$0.57}\\
(magnitude-limited) & & & & & & & \\
No weighting & \multirow{2}{*}{20.0} & \multirow{2}{*}{91.7\%} & \multirow{2}{*}{56\%}  & \multirow{2}{*}{$50.50^{+13.81}_{-1.19}$} & \multirow{2}{*}{7.50} & \multirow{2}{*}{0.15} & \multirow{2}{*}{-13.84$\pm$7.58} & \multirow{2}{*}{-1.85$\pm$1.01} \\
(magnitude-limited) & & & & & & & \\
No weighting & \multirow{2}{*}{19.0} & \multirow{2}{*}{58.9\%} & \multirow{2}{*}{24\%}  & \multirow{2}{*}{$40.0^{+10.62}_{-0.00}$} & \multirow{2}{*}{5.31} & \multirow{2}{*}{0.13} & \multirow{2}{*}{-22.24$\pm$5.24} & \multirow{2}{*}{-4.19$\pm$0.99}\\
(magnitude-limited) & & & & & & & \\
No weighting & \multirow{2}{*}{22.0} & \multirow{2}{*}{99.9\%} & \multirow{2}{*}{89\%} & \multirow{2}{*}{$71.52^{+4.82}_{-3.11}$} & \multirow{2}{*}{3.96} & \multirow{2}{*}{0.06} & \multirow{2}{*}{0.18$\pm$1.71} & \multirow{2}{*}{0.05$\pm$0.43}\\
(volume-limited) & & & & & & & \\
No weighting & \multirow{2}{*}{21.0} & \multirow{2}{*}{95.7\%} & \multirow{2}{*}{63\%} & \multirow{2}{*}{$71.52^{+2.99}_{-2.90}$} & \multirow{2}{*}{2.94} & \multirow{2}{*}{0.04} & \multirow{2}{*}{0.21$\pm$1.77} & \multirow{2}{*}{0.07$\pm$0.60} \\
(volume-limited) & & & & & & & \\
No weighting & \multirow{2}{*}{20.0} & \multirow{2}{*}{61.5\%} & \multirow{2}{*}{29\%} & \multirow{2}{*}{$71.52^{+2.37}_{-2.79}$} & \multirow{2}{*}{2.58} & \multirow{2}{*}{0.04} & \multirow{2}{*}{0.24$\pm$1.74} & \multirow{2}{*}{0.09$\pm$0.67}\\
(volume-limited) & & & & & & & \\
No weighting & \multirow{2}{*}{19.0} & \multirow{2}{*}{25.3\%} & \multirow{2}{*}{10\%} & \multirow{2}{*}{$71.52^{+3.15}_{-1.96}$} & \multirow{2}{*}{2.38} & \multirow{2}{*}{0.03} & \multirow{2}{*}{0.32$\pm$1.78} & \multirow{2}{*}{0.13$\pm$0.75}\\
(volume-limited) & & & & & & & \\
SM weighting & - & 100\% & 100\% & $70.50^{+2.40}_{-2.17}$ & 2.28 & 0.03 & 0.39$\pm$1.36 & 0.17$\pm$0.60\\
SM mismodeling by equal & - & 100\% & 100\% & $70.50^{+4.07}_{-2.65}$ & 3.36 & 0.05 & 0.75$\pm$2.00 & 0.22$\pm$0.07\\
SM mismodeling by SFR & - & 100\% & 100\% & $70.50^{+3.63}_{-2.47}$ & 3.05 & 0.04  & 0.71$\pm$1.90 & 0.23$\pm$0.62\\
equal mismodeling by SM & - & 100\% & 100\% & $71.00^{+5.21}_{-4.23}$ & 4.72 & 0.07 & 3.62$\pm$8.68 & 0.77$\pm$1.84\\
\hline\hline
 $\mathbf{d_{\rm{L,mea}} < 2500\, \rm{Mpc}}$ and SNR$>$45 - O5\\
\hline
No weighting & - & 100\% & 100\% & $70.71^{+1.47}_{-2.47}$ & 1.97 & 0.03 & $0.38\pm1.51$ & 0.19$\pm$0.77\\
No weighting ($A_{90\%}<100\, \rm{deg}^2$)& - & 100\% & 100\% & $69.90^{+6.14}_{-2.07}$ & 4.11 & 0.06 & $0.89\pm3.86$ & 0.22$\pm$0.94\\
No weighting & \multirow{2}{*}{22.0} & \multirow{2}{*}{99.9\%} & \multirow{2}{*}{89\%} & \multirow{2}{*}{$70.71^{+0.84}_{-2.86}$} & \multirow{2}{*}{1.85} & \multirow{2}{*}{0.03} & \multirow{2}{*}{$-0.14\pm1.41$} & \multirow{2}{*}{-0.08$\pm$0.76}\\
(magnitude-limited) & & & & & & & \\
No weighting & \multirow{2}{*}{21.0} & \multirow{2}{*}{95.7\%} & \multirow{2}{*}{63\%} & \multirow{2}{*}{$68.28^{+2.38}_{-1.77}$} & \multirow{2}{*}{2.08} & \multirow{2}{*}{0.03} & \multirow{2}{*}{$-3.64\pm7.46$} & \multirow{2}{*}{-1.75$\pm$3.59}\\
(magnitude-limited) & & & & & & & \\
No weighting & \multirow{2}{*}{20.0} & \multirow{2}{*}{61.5\%} & \multirow{2}{*}{29\%} & \multirow{2}{*}{$67.47^{+0.72}_{-14.36}$} & \multirow{2}{*}{7.54} & \multirow{2}{*}{0.11} & \multirow{2}{*}{$-6.97\pm7.64$} & \multirow{2}{*}{-0.92$\pm$1.01}\\
(magnitude-limited) & & & & & & & \\
    No weighting & \multirow{2}{*}{19.0} & \multirow{2}{*}{25.3\%} & \multirow{2}{*}{10\%} & \multirow{2}{*}{$44.85^{+18.77}_{-0.16}$} & \multirow{2}{*}{9.46} & \multirow{2}{*}{0.21} & \multirow{2}{*}{$-16.03\pm8.33$} & \multirow{2}{*}{-1.69$\pm$0.88}\\
(magnitude-limited) & & & & & & & \\
No weighting & \multirow{2}{*}{22.0} & \multirow{2}{*}{99.9\%} & \multirow{2}{*}{89\%} & \multirow{2}{*}{$70.71^{+1.19}_{-2.62}$} & \multirow{2}{*}{1.91} & \multirow{2}{*}{0.03} & \multirow{2}{*}{$0.22\pm1.64$} & \multirow{2}{*}{0.12$\pm$0.86}\\
(volume-limited) & & & & & & & \\
No weighting & \multirow{2}{*}{21.0} & \multirow{2}{*}{95.7\%} & \multirow{2}{*}{63\%} & \multirow{2}{*}{$70.71^{+1.26}_{-2.45}$} & \multirow{2}{*}{1.86} & \multirow{2}{*}{0.03} & \multirow{2}{*}{0.22$\pm$1.64} & \multirow{2}{*}{0.12$\pm$0.88}\\
(volume-limited) & & & & & & & \\
No weighting & \multirow{2}{*}{20.0} & \multirow{2}{*}{61.5\%} & \multirow{2}{*}{29\%} & \multirow{2}{*}{$70.71^{+1.29}_{-2.34}$} & \multirow{2}{*}{1.81} & \multirow{2}{*}{0.03} & \multirow{2}{*}{0.22$\pm$1.64} & \multirow{2}{*}{0.12$\pm$0.91}\\
(volume-limited) & & & & & & & \\
No weighting & \multirow{2}{*}{19.0} & \multirow{2}{*}{25.3\%} & \multirow{2}{*}{10\%} & \multirow{2}{*}{$70.71^{+1.23}_{-2.07}$} & \multirow{2}{*}{1.65} & \multirow{2}{*}{0.02} & \multirow{2}{*}{0.22$\pm$1.37} & \multirow{2}{*}{0.13$\pm$0.83}\\
(volume-limited) & & & & & & & \\
SM weighting & - & 100\% & 100\% & $70.71^{+0.68}_{-1.41}$ & 1.04 & 0.01 & $0.30\pm0.57$ & 0.29$\pm$0.55\\
SM mismodeling by equal & - & 100\% & 100\% & $70.71^{+2.95}_{-1.95}$ & 2.45 & 0.03 & $0.79\pm2.45$ & 0.32$\pm$1.00\\
SM mismodeling by SFR & - & 100\% & 100\% & $70.71^{+0.51}_{-2.03}$ & 1.27 & 0.02 & $-0.18\pm0.42$ & -0.14$\pm$0.33\\
equal mismodeling by SM & - & 100\% & 100\% & $69.09^{+7.06}_{-1.07}$ & 4.07 & 0.06 & $5.16\pm16.13$ & 1.27$\pm$3.96\\
\end{tabular}
\end{ruledtabular}
\label{tab:a9010deg}
\end{table*}

\subsection{Incompleteness in the volume-limited galaxy catalog}\label{vollimited}
The statistical DS analysis is based on the idea of using galaxy catalogs to infer the redshift information of GW sources. However, galaxy catalogs are magnitude-limited due to the sensitivity of telescopes. As a result, GW signals may originate from distances beyond the completeness of these catalogs, meaning the true host galaxy might not be included, or the redshift distribution of galaxies along the line of sight may not accurately represent the true distribution. This can lead to a systematic bias in the measurement of the Hubble constant (see App.\ref{app:maglimited}). In this subsection, we instead explore volume-limited galaxy catalogs and analyze the impact of catalog incompleteness on the measurement of the Hubble constant.

\begin{figure}[h!]
    \includegraphics[width=\columnwidth]{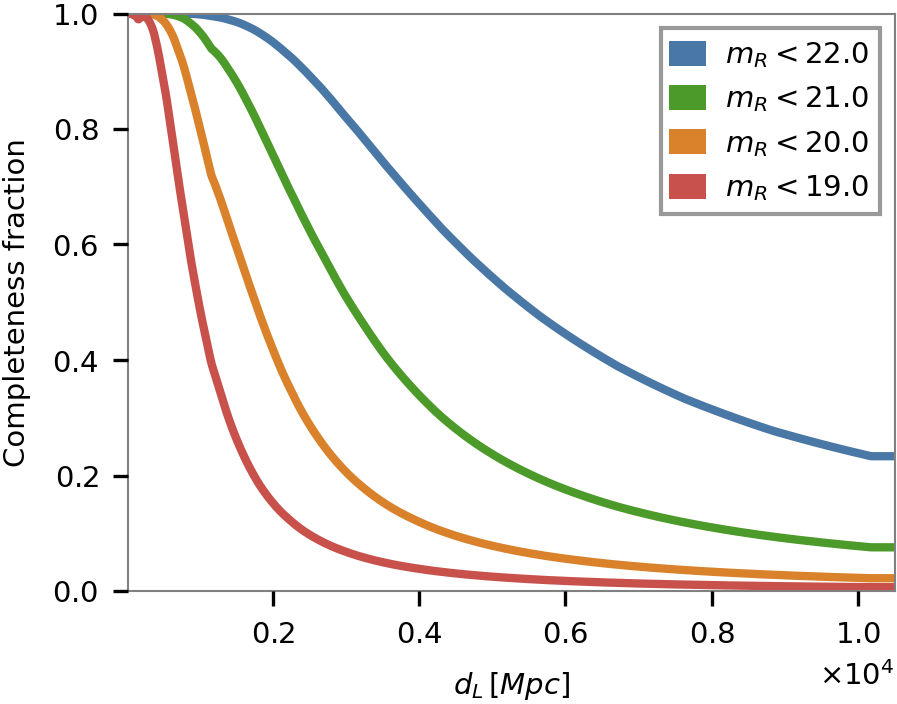}
    \caption{\texttt{\small Magneticum} catalog galaxy completeness, defined in equation \ref{eqcompleteness}, as a function of the luminosity distance. The colors represent the galaxy catalog assuming an apparent magnitude limit of 22 (blue), 21 (orange), 20 (green), and 19 (red).}
    \label{fig:completeness}
\end{figure}

Figure \ref{fig:completeness} presented the \texttt{\small Magneticum} galaxy catalog completeness for different apparent magnitude limits. As defined in \cite{Gray2020}, the galaxy catalog completeness is:
\begin{equation}
\label{eqcompleteness}
    f_{\rm completeness}\left(d_{L}, m_{\rm th}\right) = 
    \frac{
        \sum\limits_{i=1}^{N_{\rm gal}^{\rm th}\left(m_{\rm th}\right)} 
        \Theta\left(d_{L}-d_{L,i}\right)
    }{
        \sum\limits_{i=1}^{N_{\rm gal}} 
        \Theta\left(d_{L}-d_{L,i}\right)
    },
\end{equation}
where $N_{\rm gal}$ is the number of galaxies inside the \texttt{\small Magneticum} catalog, $N_{\rm gal}^{\rm th}$ is the number of galaxies after the apparent magnitude cut $m_{\rm th}$, $\Theta$ is the Heaviside function and $d_{L}$ is a reference luminosity distance. This completeness is defined as the fraction of galaxies inside the magnitude-limited catalog (with threshold $m_{\rm th}$) relative to a complete catalog (i.e., without any magnitude cut), as a function of $d_L$. As expected, the results in Fig.\ref{fig:completeness} show that the completeness fraction decreases with the reduction of the apparent magnitude limit, highlighting the impact of this effect on the intrinsic distribution of galaxies in the catalog. The abrupt change in the distribution of galaxy distances has implications for the $H_0$ posterior computed in the statistical DS analysis, as the overdensities in the galaxy distribution lead to peaks in the $H_0$ posterior. 

The incompleteness of the galaxy catalog has been extensively studied in previous works \cite{chen17, fishbach, Finke2021, Gray2020, Gray2023a, Mastrogiovanni2023, Borghi2024}, where a correction term was introduced to account for the unobserved galaxies in the catalog distribution. As pointed out in \cite{Gray2023a} and confirmed in \cite{Gray2020}, assuming a homogeneous completion for a non-uniform patch of sky can either overestimate or underestimate the galaxy contributions in certain regions. This introduces bias in the $H_0$ estimation, particularly when there is no electromagnetic data support and the contribution from the out-of-catalog term dominates. \cite{Finke2021, Gray2023a} have extended the completeness methodology to the non-uniform case in an effort to mitigate these limitations. However, this approach remains an approximation of the real data and can still be a source of bias and uncertainties. In the absence of completeness corrections, the inherent bias of a magnitude-limited galaxy catalog, which favors the brightest and/or nearest objects and excludes many distant galaxies, prevents consistency with the true redshift distribution of galaxies along the line of sight. In the context of the statistical DS methodology, this lack of distant galaxies introduces a strong bias toward lower $H_0$ values (see App.\ref{app:maglimited}), as the observed galaxy population becomes excessively concentrated at low redshifts.

One potential solution to this issue is to make the galaxy catalog volume limited. To achieve this, we follow the same steps described in previous works \cite{palmese20_sts, palmese2023}: (1) the maximum redshift for each GW event was determined by converting the higher 90\% CI bound of the luminosity distance measurement into the corresponding maximum redshift, assuming the largest value of $H_0$ in the prior; (2) we computed the absolute magnitude for the assumed apparent magnitude threshold at that maximum redshift, and then remove all galaxies within catalog that is above this absolute magnitude threshold. In all the steps, we assumed the fiducial $\Lambda$CDM cosmology. This limiting absolute magnitude represents the minimum value that a galaxy, located at the maximum redshift, must have to achieve an apparent magnitude above the limiting apparent magnitude of the galaxy sample. The limiting absolute magnitude is adjusted for each GW source, meaning that the galaxies used have different magnitude ranges (distant GW sources will have a more restrictive limiting absolute magnitude than closer GW sources in a galaxy catalog limited by a fixed apparent magnitude value) in each case. This approach relies on the assumption that the final redshift distribution of the possible host galaxies remains relatively independent of magnitude.

\begin{figure}[h!]
    \centering
    \includegraphics[width=3.44in]{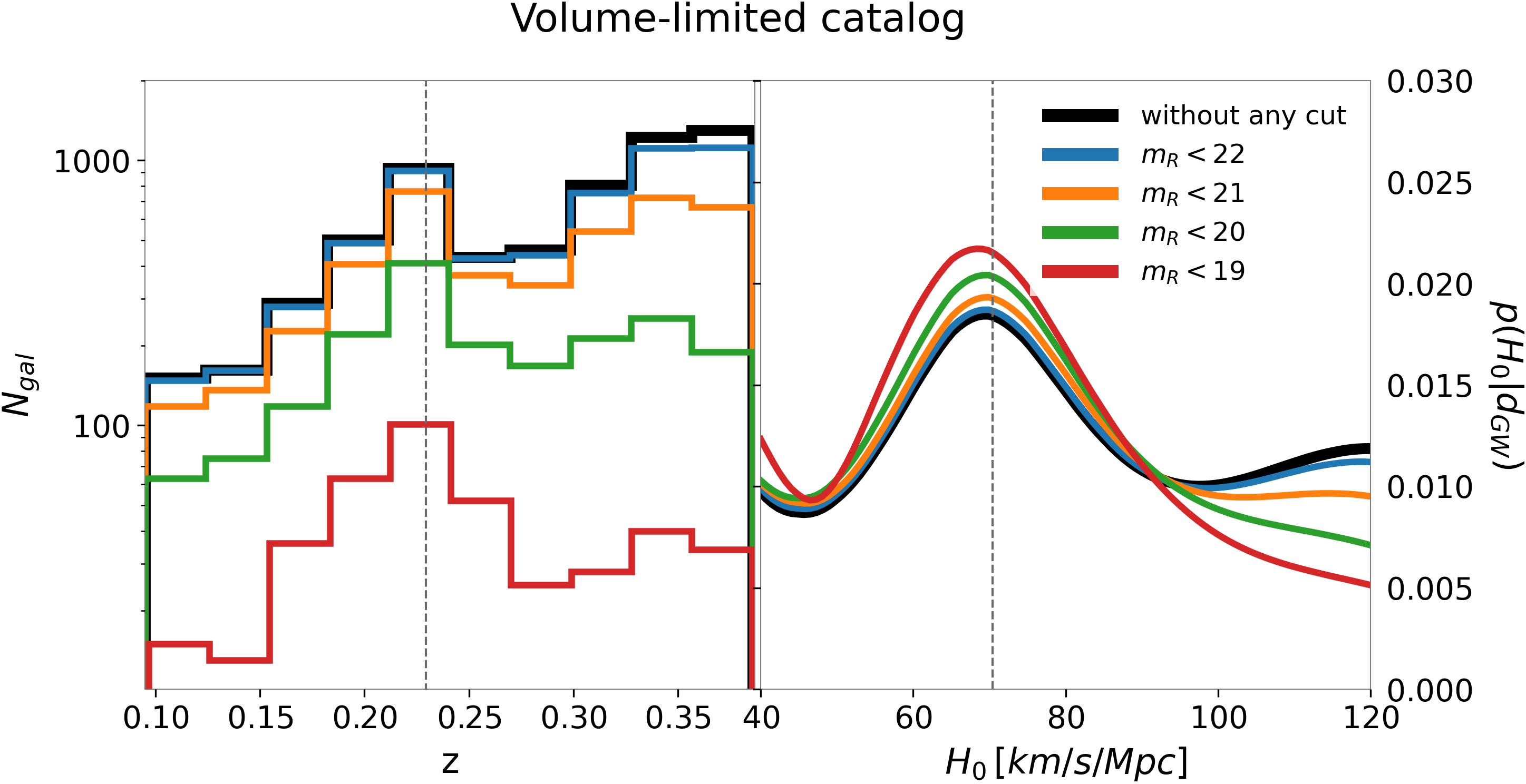}
    \caption{Redshift distribution and $H_0$ posterior for a simulated DS event using a volume-limited galaxy catalog with different magnitude thresholds. The black line shows the results using the full \texttt{\small Magneticum} galaxy catalog, while the other colors represent specific apparent magnitude cuts. The dashed gray line in the redshift distribution indicates the true redshift of the DS event, and in the $H_0$ posterior, the dashed gray line represents the fiducial value.}
    \label{fig:volumelimitedeffect}
\end{figure}

In Fig.\ref{fig:volumelimitedeffect} we present the results with different volume-limited catalogs for a simulated O4-like GW source, localized at $z \approx 0.23$ where the \texttt{\small Magneticum} catalog is already incomplete. Compared to the magnitude-limited catalog results (see Fig.\ref{fig:applimitedeffect} in App.\ref{app:maglimited}), the volume-limited case produces an $H_0$ posterior distribution more consistent with the one obtained with a complete catalog. The peak of the posterior does not shift for any absolute magnitude threshold. Nonetheless, a slight change in the $H_0$ posterior is noticeable at the boundaries. The $H_0$ posterior decreases at high $H_0$ values and slightly increases at low $H_0$ values, corresponding to the change in the redshift distribution. According to the results reported in Table\ref{tab:a9010deg}, the volume-limited catalog provides an unbiased $H_0$ measurement across all magnitude thresholds adopted in this study. In the O4 scenario, the bias reaches a minimum value of $0.18 \pm 1.71$ km/s/Mpc for $m_R<22$ and a maximum of $0.32 \pm 1.78$ km/s/Mpc for $m_R<19$. For the O5 run, the bias remains constant at $0.22$ km/s/Mpc across all apparent magnitude thresholds, with a slight decrease in the bias uncertainty for $m_{R}<19$.

\begin{figure}[h!]
    \centering
    \includegraphics[width=3.42in]{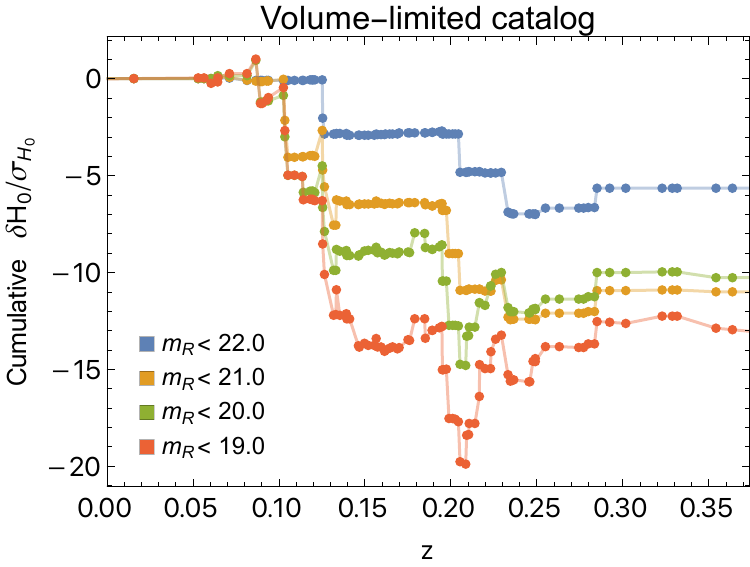}
    \caption{Cumulative sum of the ratio between the $H_0$ posterior peak offset and the standard deviation, $\delta H_0/\sigma_{H_0}$, as a function of GW source redshift for various apparent magnitude cuts (different colors). The $H_0$ posterior peak offset is defined as the difference between the $H_0$ posterior peak with both apparent and absolute magnitude cuts and that obtained without any cuts. In the text, we explain how the absolute magnitude was determined.}
    \label{fig:H0biasabsmagcut}
\end{figure}

\begin{figure*}
    \centering
    \includegraphics[width=\textwidth]{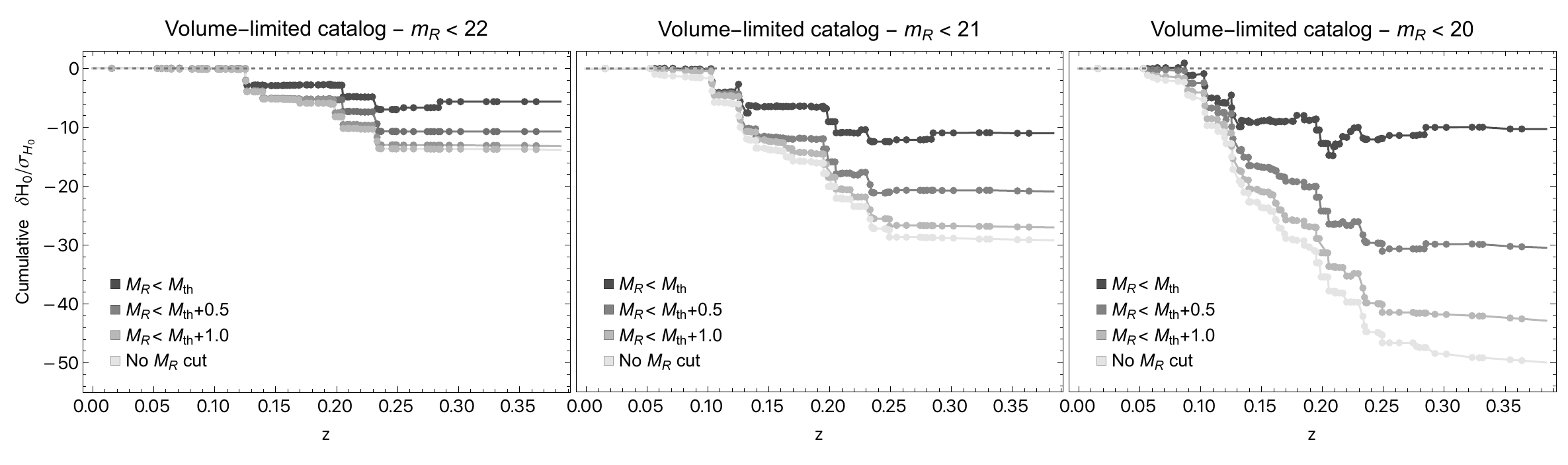}
    \caption{Cumulative sum of the ratio between the $H_0$ posterior peak offset and the standard deviation, $\delta H_0/\sigma_{H_0}$, as a function of GW source redshift for volume-limited catalogs. For each curve with a gray color scale, a distinct absolute magnitude threshold is adopted. The darkest gray represents the conservative value computed, as described in the text, while the lighter gray corresponds to a more flexible cut (higher values for the magnitude threshold), which includes fainter galaxies. 
    }
    \label{fig:H0biasmagcutflexible}
\end{figure*}

To understand the limit to which the effect of galaxy incompleteness can be controlled, Fig.\ref{fig:H0biasabsmagcut} shows the cumulative sum of the ratio between the offset of the $H_0$ posterior peak, $\delta H_0$, and the corresponding standard deviation, as a function of the GW source redshift. The offset $\delta H_0$ is defined as the difference between the $H_0$ posterior peak obtained using a catalog with an apparent (and absolute) magnitude cut and that obtained using a complete catalog, i.e., without any cut. This is computed for a volume-limited catalog using one realisation of a sample of 100 GWs. The cumulative sum value represents the total ratio of $H_0$ posterior peak offset to $\sigma_{H_{0}}$ accumulated from each GW source in the simulated sample. Each point in the figure corresponds to the sum of these ratios up to the redshift of that GW source. Fig.~\ref{fig:H0biasabsmagcut} shows that the $H_0$ offset remains less than the standard deviation ($\delta H_0/\sigma_{H_0}<1$) for GW sources detected at low redshifts ($z<0.1$). In this regime, the catalog is nearly complete, and the offset can be considered negligible. On the other hand, for GW events beyond $z=0.1$, the \texttt{\small Magneticum} catalog becomes less complete, resulting in a cumulative sum of the ratio of the $H_0$ posterior peak offset to the standard deviation of approximately -5.7, -11.0, -10.3, and -13.0 for apparent magnitude thresholds of 22, 21, 20, and 19, respectively. This represents a reduction in the ratio $\delta H_0/\sigma_{H_0}$ of approximately 59\%, 62\%, 79\%, and 85\% compared to the results obtained with the magnitude-limited catalog (see Fig.\ref{fig:H0biasappmagcut} in App.\ref{app:maglimited}) for $m_{R} < 22$, 21, 20, and 19, respectively. As can be seen in Fig. \ref{fig:H0biasabsmagcut}, the $H_0$ posterior peak offset increases with the source redshift until it reaches a plateau at high \textit{z}, where the effect of the absolute magnitude threshold is minimum compared to the apparent magnitude threshold. We demonstrate that a volume-limited catalog results in a lower cumulative sum of $\delta H_0/\sigma_{H_0}$ compared to a magnitude-limited catalog for all adopted apparent magnitude thresholds. Additionally, Fig.~\ref{fig:H0biasmagcutflexible} illustrates the transition between the volume-limited and magnitude-limited catalogs (represented by the label ``No $M_R$ cut" in the figure). 

\subsection{Mismodeling the GW galaxy host}{\label{mismodeling}}

\begin{figure}
    \includegraphics[width=\columnwidth]{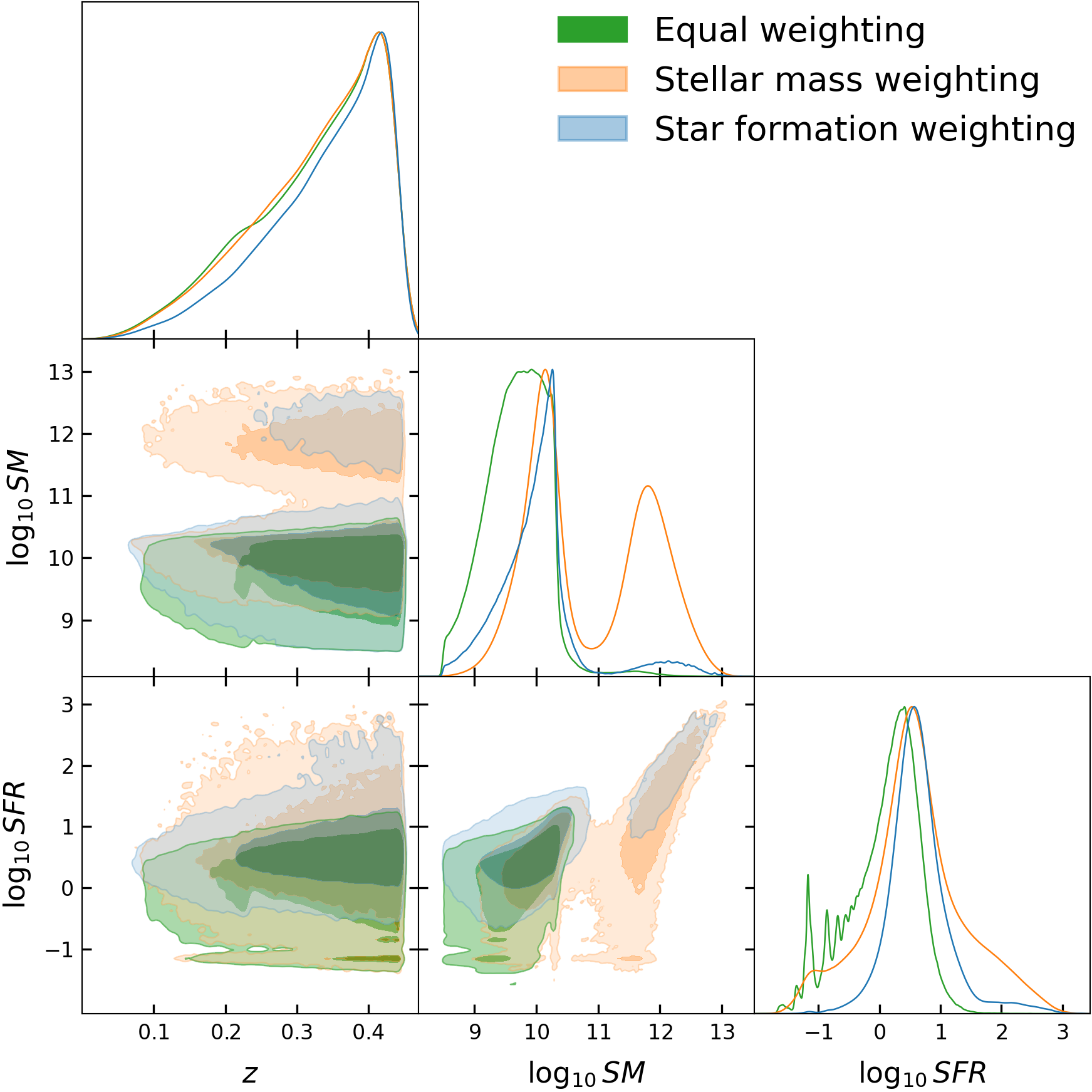}
    \caption{Distributions of stellar mass, star formation rate, and redshift from the \texttt{\small Magneticum} galaxy catalog, with each color representing results obtained using different weighting schemes.}
    \label{fig:cornerplot}
\end{figure}

We now turn our attention to exploring the potential systematic effects that may arise in the estimation of $H_0$ when incorrect weighting schemes are applied to the GW host probability. We investigate this effect using the O4-like sample of DS events, which was constructed assuming stellar mass as the GW host tracer, and investigate mismodeling by assuming equal weighting and star formation rate weighting. Figure \ref{fig:cornerplot} shows the distributions of galaxy properties (stellar mass, star formation rate, and redshift) in the \texttt{\small Magneticum} galaxy catalog for three injection sets. The 2D corner plot of SM-SFR exhibits a positive correlation between the two quantities when SFR weighting is applied, indicating that an unbiased $H_0$ estimation can still be recovered even if the SM weighting scheme is incorrectly used. However, when the true GW merger rates follow the SM, a bimodal distribution emerges in SM, breaking the positive correlation with SFR. This reveals that a systematic bias can be introduced in the $H_0$ measurement if SFR is incorrectly assumed as the GW host tracer. A similar result was found in the UniverseMachine galaxy catalog discussed in \cite{Hanselman2025}, where the bimodal distribution appeared in SFR instead of SM.

\begin{figure*}
    \centering
    \includegraphics[width=\textwidth]{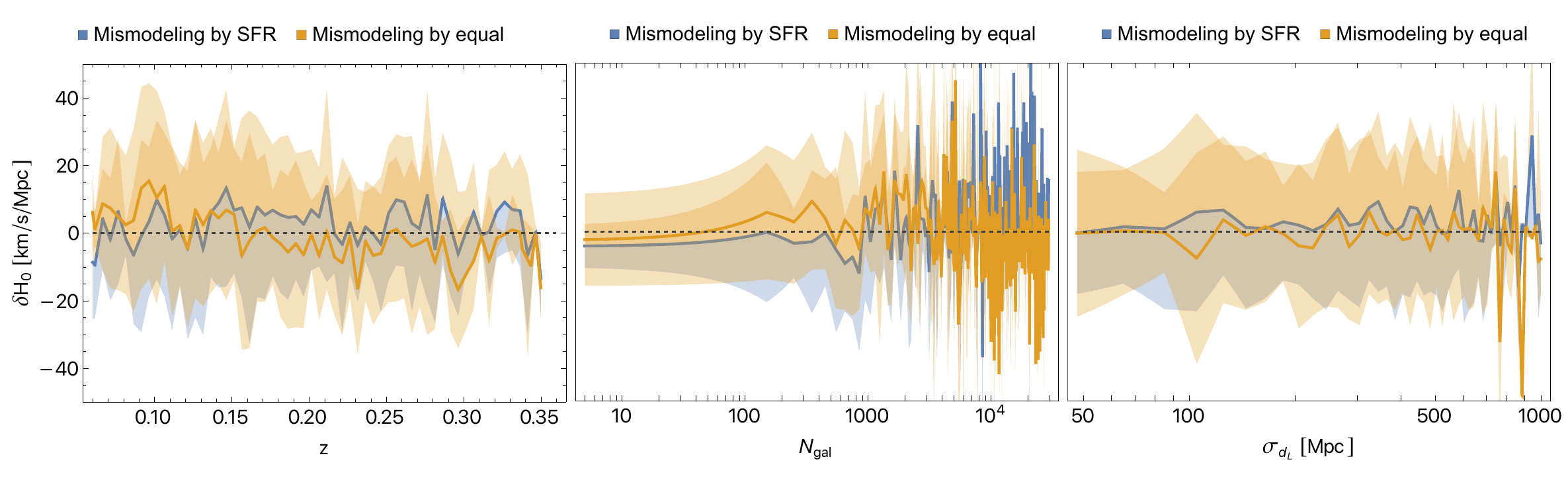}
    \caption{Offset of the Hubble constant posterior peak resulting from an incorrect weighting scheme, as a function of redshift \textit{z}, number of galaxies $N_{\rm gal}$, and luminosity distance uncertainty $\sigma_{d_{L}}$. The solid line represents the mean bias computed for each event posterior over the 30 distinct realisations, while the shaded region corresponds to the statistical uncertainty in the bias estimation. The blue color represents the results obtained under the incorrect assumption that the hosts follow the SFR, while the orange curve corresponds to the assumption of equal weights for all galaxies.}
    \label{fig:H0biasmismodeling}
\end{figure*}

The results presented in Table \ref{tab:a9010deg} indicate that an unbiased $H_0$ estimation can still be recovered (note that in the last column, the $H_0$ bias measurements are consistent with each other for the mismodeling cases) even when BBH mergers follow the SM and are mismodeled using SFR. This may seem counterintuitive based on the discussion in the previous paragraph and could lead to the incorrect conclusion that an improper weighting scheme cannot introduce bias. However, a careful analysis of the individual $H_0$ posteriors reveals that the presence of well-localized GW events (i.e., those with a low number of galaxies, $\lesssim 200$, along the line of sight) is responsible for producing an unbiased estimation in the combined result. Figure \ref{fig:H0biasmismodeling} shows how the mean of the $H_0$ posterior peak offset varies with redshift, the number of possible host galaxies ($N_{\rm gal}$), and the uncertainty in the luminosity distance of the GW source. The $H_0$ offset is defined as the difference between the posterior peak obtained using an incorrect weighting scheme and that obtained using the correct one. We used bin widths of 0.005 for $z$, 100 for $N_{\rm gal}$, and 20 Mpc for $\sigma_{d_L}$, respectively. Assuming an incorrect GW host weighting scheme, we observe that the individual $H_0$ posterior peak can be offset by up to $\delta H_0 \approx 40$ km/s/Mpc, or even higher, in the case of imprecise luminosity distance and sky localisation measurements (which are related to $N_{\rm gal}$). Our results are in agreement with those of \cite{Hanselman2025}, which demonstrated that an incorrect weighting scheme can lead to a significant bias in the ``transitional regime" when $N_{\rm gal} = 10^{2}–10^{4}$.

\begin{figure}
    \includegraphics[width=\columnwidth]{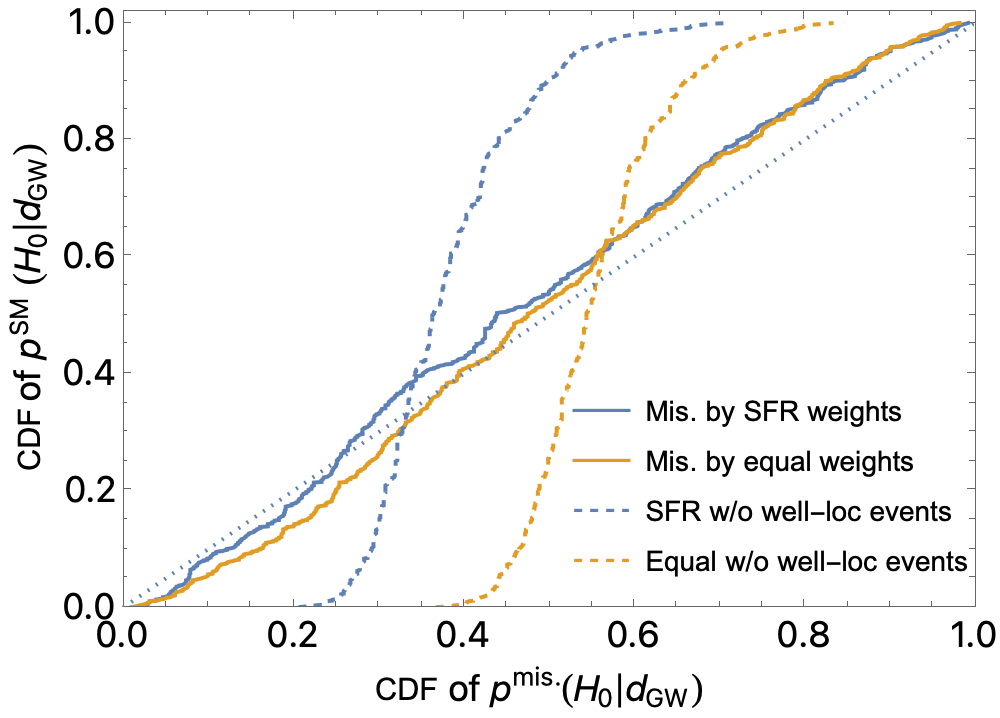}
    \caption{PP plot comparing the posterior distributions of $H_0$ under different weighting schemes: the corrected scheme (SM) versus the incorrect ones (SFR or equal weights). The y-axis shows the cumulative distribution function (CDF) of the posterior found with the corrected weighting scheme, and the x-axis shows the CDF of the posterior assuming the incorrect weighting scheme. The solid lines represent the results considering all 100 GW events in the sample, while dashed lines correspond to the case where well-localized GW events are excluded. Deviations from the identity line (dotted) indicate discrepancies between the posteriors caused by incorrect modeling of the weighting scheme.}
    \label{fig:ppplot}
\end{figure}

In addition, we explored the results discussed above using probability-probability (PP) plots \cite{Wilk1968} for the $H_0$ posterior, assuming the correct weighting scheme (stellar mass as a GW host tracer) and an incorrect weighting scheme (weighting by SFR or equal probability). The PP plot presents the cumulative distributions of two different datasets; if they follow the identity line in the range $\left[0,1\right]$, the two distributions are approximately equal, indicating no bias. Using the mean of $H_0$ posterior from 30 independent realisations, each considering 100 detected GW events with $A_{90\%}<10\, \rm{deg}^2$, we constructed the PP plot (see solid lines in Fig.\ref{fig:ppplot} for the two mismodel case). In both cases, we observe a slight deviation of the curves from the identity line. Additionally, we repeated the analysis presented in Table \ref{tab:a9010deg}, but excluding all well-localized GW events. These results are represented as a dashed line in Fig.\ref{fig:ppplot}. The interpretation of this PP plot shows that the absence of well-localized events leads to an offset of the $H_0$ posterior peak toward higher values and leads to a broader posterior distribution. In particular, when the star formation rate (or an equal-weighting scheme) is incorrectly applied, the estimated values of the Hubble constant are $H_0 = 74.5^{+12.2}_{-12.5}$ km/s/Mpc ($H_0 = 71.0^{+12.3}_{-16.9}$ km/s/Mpc).
 
\subsection{Understand the $H_0$ precision from the dark sirens: the impact of selection cuts}
The quality of cosmological constraints derived from DSs is inherently tied to the coverage and completeness of the galaxy catalog, as well as the precision of the gravitational-wave luminosity distance measurement. Therefore, identifying and selecting the most informative DSs is essential for optimizing their contribution to constraining cosmological parameters. In the following paragraph, we investigate how the $H_0$ forecasts change by different selection criteria, including (1) the uncertainty in the luminosity distance, (2) varying redshift ranges, and (3) the sky localisation area.

First, we report constraints on $H_0$ constraints derived from a DS mock catalog, assuming an equal-weight scheme with different selection criteria based on the luminosity distance measurement of GW sources. Specifically, we consider three distinct cuts on the fractional error in the GW luminosity distance measurement, defined as the ratio of the uncertainty to the mean value ($\sigma_{d_L}/ \overline{d}_{L}$): $<0.2, 0.15$, and $0.1$. 
All selected DSs satisfied the sky localisation criterion $A_{90\%}<10\,$ deg$^2$. In general, we find that events with a fractional error of 20\% can still contribute to improving constraints on $H_0$, leading to a reduction of 18\% and 8\% in the uncertainty compared to the cases with  $\sigma_{d_L}/ \overline{d}_{L} =$ 0.1 and 0.15, respectively. In addition, we find that with a sample of approximately 1,440 DSs detected by LVK O4, it is possible to achieve an $H_0$ uncertainty of 0.65 km/s/Mpc. These results are summarized in Table \ref{tab:selcuts}.

Furthermore, we explore how our forecasts change for deeper redshift ranges and investigate the limits of statistical DS analysis as a cosmological probe, achievable with current GW detections. We found that the most informative DSs are localized at low redshift ($z<0.1$), and the $H_0$ uncertainty decreases only by 0.1 km/s/Mpc when including sources up to $z < 0.3$. This result arises from the significant degradation in the 3D localisation accuracy of events at high redshift, which makes it difficult to identify their host galaxies, resulting in uninformative $H_0$ posteriors. In the next-generation GW detectors, the sky localisation volume will be measured with greater precision, enabling the extraction of more cosmological information from high-redshift GW sources. This scenario was explored by \cite{Muttoni2023} in the era of third-generation gravitational wave detectors, such as the Einstein Telescope and Cosmic Explorer, with a complete galaxy catalog extending to $z < 1$. The authors demonstrated that with a sample of well-localized DSs (SNR $> 300$), it is possible to achieve a precision of 0.8\% for the Hubble constant and 10\% for the matter energy density parameter.

Finally, we also investigate how the precision of $H_0$ degrades as the GW sky localisation area increases. This analysis is particularly important for understanding the selection criteria that ensure the identification of the most cosmologically informative DS events. Specifically for this analysis, we selected DS events that satisfy the sky localization criterion $A_{90\%} < 100\, \mathrm{deg}^2$. In the previous paragraph, we discussed how the precision of $H_0$ varies as a function of the luminosity distance uncertainty. We observed that as the luminosity distance uncertainty increases, we approach the non-informative limit, where DS events are unable to improve the precision of $H_0$. This same limit can also be investigated in terms of sky localisation area, as larger areas contain a greater number of galaxies, making it more challenging to identify the host galaxy. Consequently, we can assess the impact on $H_0$ measurements in terms of catalog coverage over the localisation area or the redshift depth. We found that it is possible to identify informative DSs with sky localisation areas in the range $ 50\, \rm{deg}^2 < A_{90\%} < 100\, \rm{deg}^2$ (see the results reported in Table~\ref{tab:selcuts}). These events provide a reduction of approximately $\sim 10\%(80\%)$ in the $H_0$ uncertainty compared to the results for $A_{90\%} < 50\, \rm{deg}^2$ ($A_{90\%} < 10\, \rm{deg}^2$). The most significant improvement, a reduction of approximately $\sim 60\%$ in the $H_0$ precision, occurs with the inclusion of events in the range $20\, \rm{deg}^2 < A_{90\%} < 50\, \rm{deg}^2$. This improvement degrades to approximately 10\% when including events in the range $50\, \rm{deg}^2 < A_{90\%} < 100\, \rm{deg}^2$.

\begin{table}
\caption{$H_0$ measurements for different selection cuts (redshift range, luminosity distance uncertainty, and sky localisation area) applied to GW detections with $A_{90\%}<10\, \rm{deg}^2$ (except where explicitly stated otherwise) and a luminosity distance thresholds of $1600\, \rm{Mpc}$.}
\begin{ruledtabular}
\begin{tabular}{cccc}
 Selection cuts & $N_{\rm GW}$ & $H_0$ [km/s/Mpc] & $\sigma_{H_{0}}$ [km/s/Mpc] \\
 \hline\hline
$z<0.1$ & 506 & $70.71^{+0.88}_{-0.31}$ & 0.6 \\
$z<0.2$ & 1951 & $71.52^{+0.35}_{-0.84}$ & 0.6\\
$z<0.3$ & 2834 & $70.71^{+0.48}_{-0.52}$ & 0.5 \\
$\sigma_{d_{L}}/\overline{d}_{L}<0.1$ & 482 & $71.52^{+0.62}_{-0.96}$ & 0.79 \\
$\sigma_{d_{L}}/\overline{d}_{L}<0.15$ & 897 & $71.51^{+0.53}_{-0.88}$ & 0.71 \\
$\sigma_{d_{L}}/\overline{d}_{L}<0.2$ & 1339 & $70.71^{+0.92}_{-0.39}$ & 0.65 \\
$A_{90\%}<10\,\rm{deg}^2$ & 314 & $69.90^{+3.39}_{-3.20}$ & 3.29 \\
$A_{90\%}<20\,\rm{deg}^2$ & 660 & $69.90^{+1.23}_{-2.42}$ & 1.83 \\
$A_{90\%}<50\,\rm{deg}^2$ & 1160 & $69.90^{+0.60}_{-0.86}$ & 0.73 \\
$A_{90\%}<100\,\rm{deg}^2$ & 1630 & $69.90^{+0.82}_{-0.49}$ & 0.66 \\
\end{tabular}
\end{ruledtabular}
\label{tab:selcuts}
\end{table}

\section{Conclusions}\label{conclusion}
The advent of dedicated deep-field surveys, combined with the rapidly increasing number of gravitational wave detections, is set to establish DSs as a powerful tool for cosmological probing. Therefore, given the approximation of the percent-level precision, it is essential to carefully consider the potential systematic biases in $H_0$ inference that could arise in such analyses. Here we investigate the systematic of the DSs analysis when (1) incorrectly weighted the GW host, and (2) an incomplete galaxy catalog is used. We also investigated the potential of DSs for cosmological parameter inference.  

First, we performed a series of end-to-end simulations of binary black hole mergers from the LVK detector network with a sensitivity similar to O4 and O5 runs, combined with a galaxy mock catalog generated from the \texttt{\small Magneticum} simulations. We found it possible to obtain unbiased measurements when recovering with the same injected model. The average results over the 30 realisation provided an initial indication that it is possible to recover unbiased $H_0$ measurements even when injected by SM and incorrectly weighting the host galaxy scheme assuming SFR or equal probability. However, a detailed analysis of the individual posteriors revealed that the $H_0$ posterior peak can be offset by approximately $\delta H_0 \gtrsim 40$ km/s/Mpc for GW detections with $\sigma_{d_L}>10^{2}\, \rm{Mpc}$ and $N_{\rm gal}>10^{3}$. This result was masked by the presence of well-localized events, which did not exhibit significant bias in the $H_0$ measurements. These results highlight a potential solution to reduce such biases, as the next generation of GW facilities will improve the localisation volume of gravitational waves, leading to an increase in the number of well-localized events observed. 

The absence of incompleteness correction in a volume-limited galaxy catalog does not introduce any bias in the $H_0$ measurement. We obtained $H_0$ measurements that are statistically consistent within 1$\sigma$ for all adopted apparent magnitude thresholds. These results demonstrate the capability of the volume-limited catalog to accurately recover the true distribution of galaxy overdensities and underdensities along the line of sight.

Finally, we characterized the constraining power of DSs. We find a $H_0$ precision of 6\%(3\%) for 100 O4-like (O5-like) detected GW events with a 90\% CI sky localisation area less than 10 deg$^2$, assuming no host-galaxy weighting. This number of detections is expected to be achieved after approximately 8 years of observation at O4 sensitivity and 3 years at O5 sensitivity with the LVK. The precision decreased to 13\%~(6\%) for the O4~(O5) run when the sky localisation area increased to $<100$ deg$^2$. We show that weighting the potential host galaxies by stellar mass improves the constraints, reducing the uncertainty by approximately half compared to the case without weighting for both O4-like and O5-like configurations. Specifically, the uncertainty decreases by 2.28 km/s/Mpc in the O4-like configuration and by 1.04 km/s/Mpc in the O5-like configuration, resulting in an $H_0$ precision of 3\%(1\%) for O4(O5) run. A detailed analysis on how the selection criteria affect the $H_0$ forecast reveals that events with $0.15 < \sigma_{d_{L}}/\overline{d}_{L} < 0.2$ and $50\,\rm{deg}^2<A_{90\%}<100\,\rm{deg}^2$ can modestly improve the $H_0$ measurement. The most informative events are localized at low redshift, but DSs at $z>0.2$ lead to a decrease of 0.1 km/s/Mpc in the $H_0$ uncertainty.

The results presented here are based on the simplifying assumption that galaxy redshifts are measured perfectly, i.e., redshift uncertainties are ignored. This is an optimistic scenario, as galaxies are likely to have significant photometric redshift uncertainties due to the lack of spectroscopic data at the redshift limits considered in this study. As demonstrated in \cite{Gair2023}, photometric redshift uncertainties broaden the $H_0$ posterior and, if an inconsistent Bayesian analysis is adopted, can introduce bias. These results were derived under the assumption of a Gaussian approximation for the redshift likelihood. However, in real-world applications, machine learning or template fitting algorithms are typically used to compute the photo-$z$ probability distribution function. It is therefore crucial to investigate how non-Gaussian features in the photo-$z$ PDF impact the results of this analysis. We leave this investigation to a future study.

\section*{Acknowledgements}
We acknowledge Alex Saro and Klaus Dolag for the \texttt{Magneticum} \citep{Hirschmann2014}. The authors made use of Sci-Mind servers machines developed by the CBPF AI LAB team and would like to thank Paulo Russano and Marcelo Portes de Albuquerque for all the support in infrastructure matters. VA thanks Antonella Palmese for her hospitality and useful discussions during a visit to McWilliams Center for Cosmology \& Astrophysics. CRB acknowledges the financial support from CNPq (316072/2021-4) and from FAPERJ (grants 201.456/2022 and 210.330/2022) and the FINEP contract 01.22.0505.00 (ref. 1891/22). 
\nocite{*}

\appendix

\section{Incompleteness galaxy catalog: magnitude-limited case}\label{app:maglimited}
In this appendix, we present a complementary discussion about galaxy incompleteness in Hubble constant measurements, as introduced in Section \ref{vollimited}, within the context of magnitude-limited galaxy surveys.

\begin{figure}
    \centering
    \includegraphics[width=3.45in]{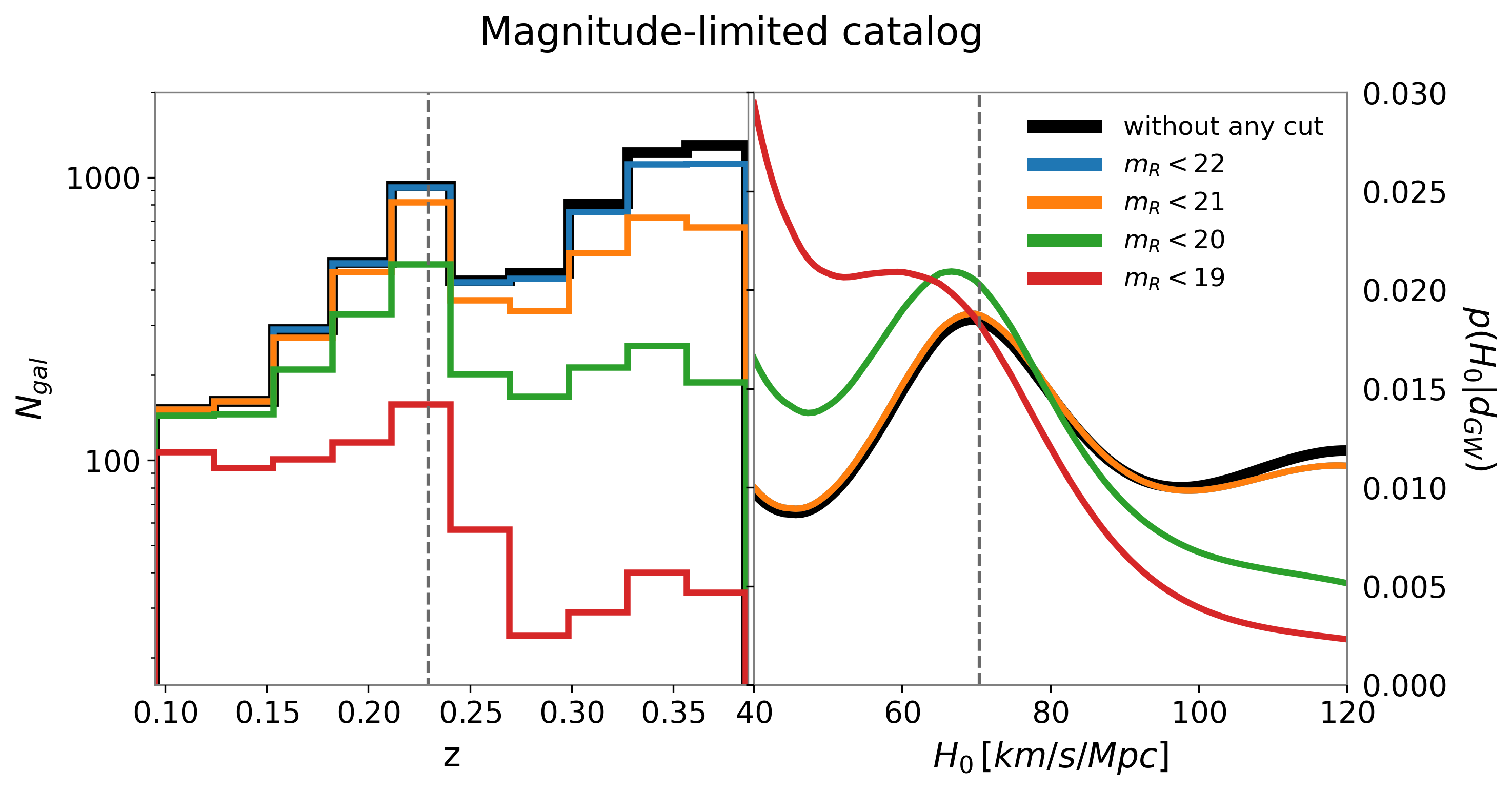}
    \caption{Effect of the apparent magnitude cut on the redshift distribution and the $H_0$ posterior for a simulated DS event. The black curve corresponds to results derived from the complete \texttt{\small Magneticum} galaxy catalog, while the colored curves denote subsets defined by specific apparent magnitude limits. In the redshift distribution panel, the dashed gray line marks the true redshift of the DS event; in the $H_0$ posterior panel, it indicates the fiducial value.}
    \label{fig:applimitedeffect}
\end{figure}

To exemplify the effect of magnitude-limited galaxy incompleteness, in the left panel of Fig.\ref{fig:applimitedeffect} we applied various apparent magnitude limits on the \textit{r}-band for a simulated O4-like GW source, localized at $z \approx 0.23$ where the \texttt{\small Magneticum} catalog is already incomplete. As the catalogs become less complete, fewer galaxies are visible at high redshift ($z \approx 0.23$), which reduces the peak of the redshift distribution, with the majority of galaxies concentrated at lower \textit{z}. This change leads to a systematic bias in the $H_0$ posterior: the maximum of the posterior moves to lower $H_0$ values.

The results presented in Table~\ref{tab:a9010deg} for the magnitude-limited surveys indicate that the final $H_0$ constraint remains unbiased when the galaxy catalog is incomplete, with a minimum apparent magnitude threshold of 21.0, corresponding to 86\% completeness for the O4 run configuration, and a threshold of 20.0 (29\% completeness) for the O5-like scenario. A more restrictive apparent magnitude threshold, $m_{R}<19.0$, will result in a biased $H_0$ measurement, with maximum bias of -22.24 km/s/Mpc and -16.03 km/s/Mpc for the O4 and O5 observing runs, respectively. The difference between the results is mainly due to the fact that we are considering a higher SNR threshold for O5, resulting in events with better sky localisation and, consequently, less susceptibility to the effect of catalog incompleteness.

\begin{figure}
    \centering
    \includegraphics[width=3.42in]{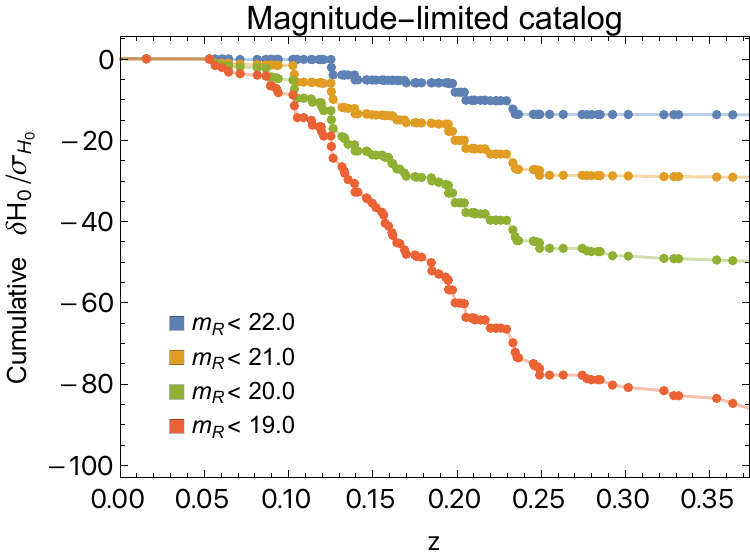}
    \caption{Cumulative distribution of $\delta H_0/\sigma_{H_0}$, the offset of the $H_0$ posterior peak normalized by its standard deviation, as a function of GW source redshift, for different apparent magnitude limits (shown in color). This corresponds to the magnitude-limited survey version of Fig.~\ref{fig:H0biasappmagcut}.}
    \label{fig:H0biasappmagcut}
\end{figure}

Fig.~\ref{fig:H0biasappmagcut} demonstrates that the $H_0$ offset is negligible ($\delta H_0/\sigma_{H_0}<1$) for nearby GW sources ($z\lesssim0.06$), where the galaxy catalog is nearly complete. Beyond this limit, the cumulative value of the normalized $H_0$ offset begins to increase, a regime increasingly affected by catalog incompleteness. For an apparent magnitude limit of 19, we find that the cumulative sum of $\delta H_{0}/\sigma_{H_{0}}$ reaches approximately -84.8, highlighting the potential for significant bias under strong incompleteness.
As expected, an apparent magnitude threshold of 22 yields the minimum cumulative sum value, -13.8. We can also observe an abrupt increase in the offset for $m_R < 20$, which is a direct consequence of the decrease in the galaxy completeness fraction within the distance range of interest (see Fig.\ref{fig:completeness}).

\bibliography{apssamp}

\end{document}